%% file: 0tmtt2022.tex
\documentclass[journal]{IEEEtran}
\usepackage{cite}
\ifCLASSINFOpdf
\else
\fi
\usepackage{amsmath}% for double integral symbol in this template
\usepackage{graphicx}% for figures
\usepackage{multirow}% to allow multiple-row elements in tabular environment
\usepackage[none]{hyphenat}% turn off hyphenation to make text extraction and indexing easier
\usepackage{float}% better control of floating figures and tables
\usepackage{subfig}% for subfigures within figures
\usepackage{dblfloatfix}% fix to allow page-wide floats at bottom of page
\usepackage{color}
\usepackage{cite}
\usepackage[section]{placeins}
\usepackage{perpage}
\MakeSorted{figure}
\MakeSorted{table}

\usepackage{t1enc}% allows access to various special characters
\usepackage{times}% change font to Nimbus Roman (based on Times Roman)
\begin{document}
%
% paper title
% Titles are generally capitalized except for words such as a, an, and, as,
% at, but, by, for, in, nor, of, on, or, the, to and up, which are usually
% not capitalized unless they are the first or last word of the title.
% Linebreaks \\ can be used within to get better formatting as desired.
% Do not put math or special symbols in the title.
\title{Instinctual Interference Adaptation in a Low-Power High-Linearity Receiver With Combined Feedforward and Feedback Control}
\title{Sub-1ms Instinctual Interference Adaptation in a Low-Power High-Linearity Receiver With Feedforward and Feedback Control Systems (Need Change}
\title{Sub-1ms Instinctual Interference Adaptive GaN LNA Front-End with Power and Linearity Tuning }
% \title{??????}

%
%
% author names and IEEE memberships
% note positions of commas and nonbreaking spaces ( ~ ) LaTeX will not break
% a structure at a ~ so this keeps an author's name from being broken across
% two lines.
% use \thanks{} to gain access to the first footnote area
% a separate \thanks must be used for each paragraph as LaTeX2e's \thanks
% was not built to handle multiple paragraphs
%

%%%%%%%%%%%%%%%AUTHOR%%%%%%%%%%%%%%%%%%%%%%%%%%%%%%
% \author{Michael~Shell,~\IEEEmembership{Member,~IEEE,}
%         John~Doe,~\IEEEmembership{Fellow,~OSA,}
%         and~Jane~Doe,~\IEEEmembership{Life~Fellow,~IEEE}% <-this % stops a space
% \thanks{M. Shell was with the Department
% of Electrical and Computer Engineering, Georgia Institute of Technology, Atlanta,
% GA, 30332 USA e-mail: (see http://www.michaelshell.org/contact.html).}% <-this % stops a space
% \thanks{J. Doe and J. Doe are with Anonymous University.}% <-this % stops a space
% \thanks{Manuscript received April 19, 2005; revised August 26, 2015.}}

\author{Jie~Yang,~\IEEEmembership{Graduate Student Member,~IEEE,}
        Baibhab~Chatterjee,~\IEEEmembership{Member,~IEEE,}
        Mohammad~Abu~Khater,~\IEEEmembership{Member,~IEEE,}
        Mattias~Thorsell,~\IEEEmembership{Member,~IEEE,}
        Sten~E.~Gunnarsson,~\IEEEmembership{Senior Member,~IEEE,}
        Tero~Kiuru,
        and~Shreyas~Sen,~\IEEEmembership{Senior Member,~IEEE}% <-this % stops a space
\thanks{This work is sponsored by SAAB AB, Sweden, and SAAB Inc., USA.}
\thanks{J. Yang, B. Chatterjee, M. Abu Khater, and S. Sen are with the Department of Electrical and Computer Engineering, Purdue University, West Lafayette, Indiana, 47907 USA e-mail: (yang1122@purdue.edu, shreyas@purdue.edu).}% <-this % stops a space
\thanks{M. Thorsell, S. E. Gunnarsson and T. Kiuru are with SAAB AB, Sweden (email: mattias.thorsell@saabgroup.com).}
\thanks{B. Edward is with SAAB Inc, USA (email: Brian.Edward@saabinc.com).}
\thanks{This paper is an extended version of the MWCL paper \cite{imspaper}.}
\vspace{-1em}}

\maketitle

% As a general rule, do not put math, special symbols or citations
% in the abstract or keywords.

\begin{abstract}
One of the major challenges in communication, radar, and electronic warfare receivers arises from nearby device interference. The paper presents a 2-6 GHz GaN LNA front-end with onboard sensing, processing, and feedback utilizing microcontroller-based controls to achieve adaptation to a variety of interference scenarios through power and linearity regulations. The utilization of GaN LNA provides high power handling capability (30 dBm) and high linearity (OIP3= 30 dBm) for radar and EW applications. The system permits an LNA power consumption to tune from 500 mW to 2 W (4X increase) in order to adjust the linearity from P\textsubscript{1dB,IN}=-10.5 dBm to 0.5 dBm (>10X increase). Across the tuning range, the noise figure increases by approximately 0.4 dB. Feedback control methods are presented with backgrounds from control theory. The rest of the controls consume $\leq$10$\%$ (100 mW) of nominal LNA power (1 W) to achieve an adaptation time <1 ms.

% In the low power mode, the LNA consumes about 500 mW of power to achieve a P\textsubscript{1dB,IN} of -10.5 dBm; in the high power mode, the LNA drains about 2 W of power to achieve a P\textsubscript{1dB,IN} of 0.5 dBm. The P\textsubscript{1dB,IN} increased by 11 dB (> 10X) with 4X increase in power. The adaptation time for the LNA to become linear with the presence of a blocker is $<$ 1 ms. 

% Two receivers are designed with one involving both feedforward and feedback control for linearity adaptation and the other involving only a feedback control. Both receivers are populated with commercial off-the-shelf (COTS) components to adapt with a 2-6 GHz GaN low-noise amplifier (LNA). Three control methods: incremental adaptation only, look-up table (LUT), and one-shot + incremental adaptation are compared.

% Other control loop device in the feedback only control consumes $\approx$ 55 mW of power.

\end{abstract}

% Note that keywords are not normally used for peerreview papers.

% For peer review papers, you can put extra information on the cover
% page as needed:
% \ifCLASSOPTIONpeerreview
% \begin{center} \bfseries EDICS Category: 3-BBND \end{center}
% \fi
%
% For peerreview papers, this IEEEtran command inserts a page break and
% creates the second title. It will be ignored for other modes.
\IEEEpeerreviewmaketitle

\vspace{-1em}
\section{Introduction}

% What is needed (A conceptual picture). 3 things 1) Tunable LNA 2) Built-in Sensors 3) Control Logic

% With continuing technology advances and rising number of technologies, interference is more prevalent in our day to day life. The interference especially a blocker is causing the receiver system to saturate and thus an undecodable signal. The first step to combat a blocker is to have a low-noise amplifier (LNA) with a high OIP3; to achieve that, GaN LNAs are favored with the ability to have a higher power handeling capacity and linearity. Many of the LNAs today are designed to withstand the worse case interferer. The downfall to the highly linear LNA is the constant higher power consumption. This calls for a tunable receiver with a tunable LNA to change the linearity and thus changes the power consumption when necessary.

With continuous advances in communication, radar, and electronic warfare (EW) technologies, the receivers (Rx) are becoming more susceptible to nearby interferences. Unlike the intentional jamming that deliberately saturates the receiver system to produce an unusable signal, unintentional interferences are more prevalent such as self-interference, and adjacent channel interference and reflection by neighboring devices\cite{interference1,EW}. To combat unintentional interferences, some of today's receivers are designed to operate in the worst-case condition at the cost of extra power consumption. When the radar and EW receivers are implemented in large arrays, the extra power consumption can quickly add up and may require extra cooling. The increasing interferences caused extra power consumption calls for an adaptive interference-tolerant low-power RF Rx system.

The first step to the adaptive interference-tolerant low-power RF receiver system especially in the radar and EW receivers is the ability to sustain a high input level. GaN LNAs have been widely used due to their higher power handling capabilities (>30 dBm) than traditional GaAs or CMOS LNAs (20 dBm). Traditional LNAs usually require the implementation of a limiter in the input path for protection from high input power while the GaN LNA is able to function standalone \cite{EW}. The next step is the ability to adjust the receiver power consumption according to the input power as opposed to continuous operation under worse-case conditions. Operation of the receiver system in low-power mode can significantly reduce the power consumption and the need for cooling when the input is small in a large radar and EW receiver array. 

% To supplement the tunable LNA, built-in sensors and necessary control logic are required for a more complete receiver. 

% As shown in Fig. \ref{fig:concept}, with adaptation, the system is able to be less gain compressed as well as achieving a higher 3rd order intermodulation compression.

% One two column Fig. at the start of Pg 2 covering background. 

% A 1/2 to 1-page survey of previous work

% including my papers (system level), orthogonal tunability and columbia papers (), others (?) 

\subsection{Background of Adaptive Receivers and Related Works}
\begin{figure}
\centering
\includegraphics[width=85mm]{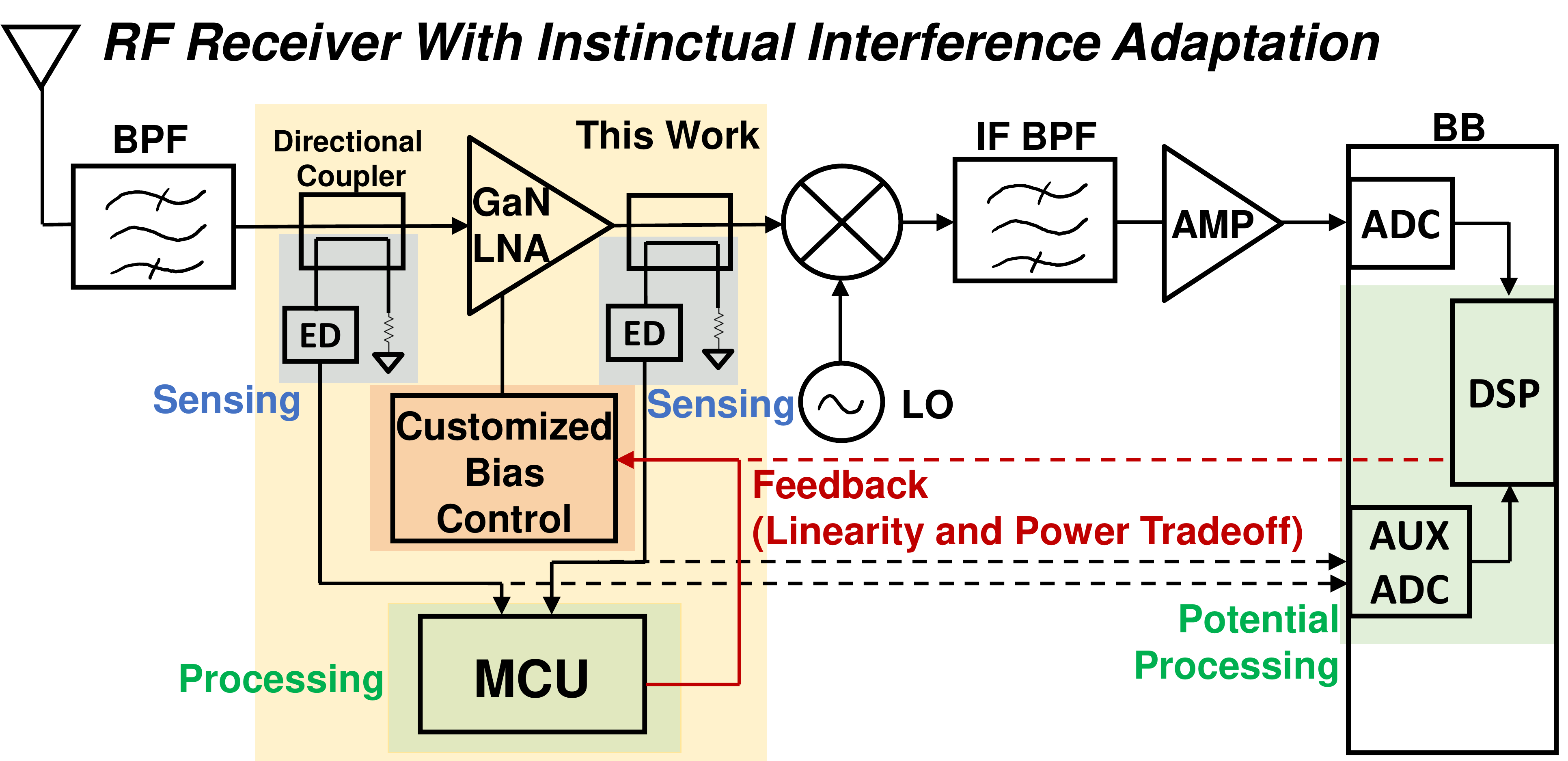}
\caption{Block Diagram for RF Receiver with Instinctual Interference Adaptation.}
% When the interference is large as compared to the signal, the control loop helps in increasing the linearity. When multiple blockers are present, the control loop helps in reducing the in-band IM3 component.
\vspace{-1.5em}
\label{fig:concept}
\end{figure}

In the effort of achieving a low-power and interference adaptive Rx, some previous works have focused on linearizing mixers using frequency translation by compressing third-order intermodulation product (IM3)  \cite{linear1,linear2,linear3}; however, if the Low-Noise Amplifiers (LNA) are already saturated, linearizing the later stages gains little advantage. Other works contributed on creating ultra-low-power LNA using the current reuse and forward body biasing techniques \cite{lowpowerLNA1,lowpowerLNA2,lowpower1}, variable-gain LNA \cite{tunableLNA2, tunableLNA4, gain1,gain2}, an orthogonally tunable LNA where input third-order intercept point (IIP3) and gain can be individually changed through the bias tuning knobs \cite{tunableLNA1}, or a combination of low-power and variable-gain LNA \cite{tunableLNA3}. The variable gain and IIP3 in the design of LNA can assist the receiver to become more interference tolerant in terms of both large signal saturation and small signal nonlinearity; however, these works only included circuit level designs without a complete system design. With the purpose of developing an adaptive receiver, dynamic bias (tuning of the gate and drain voltage) for the optimization of the signal-to-noise and distortion (SNDR) for a Gallium-Nitride (GaN) LNA is explored in \cite{SNDR}; however, the system-level considerations for the dynamic bias technique were not included.

In \cite{rxsystem}, the authors utilized the orthogonally tunable LNA to implement a use-aware adaptive RF transceiver system where different low-power adaptation modes are designed for different throughput requirements but disregard the tuning speed of the Rx. Other implementations of an adaptive RF communication system use error vector magnitude (EVM) and look up table (LUT) with stored tuning conditions for the LNA and mixer to optimally trade off power and performance \cite{vizor}. Advancements were made later by considering the process variation of the components with tuning adjustments included in the design process \cite{provizor}. These contributions provide a thorough design of the Rx system, but most of the results are still simulation-based, and utilization of only the LUT does not provide local feedback for further tuning of the system.  

Other adaptive systems with implementation of COTS, IC, or simulation can be found in \cite{comp1,comp2,comp3,comp4,comp5,comp7,comp9,comp10}.

\vspace{-1em}
\subsection{Proposed solution}

This paper presents the first building block, a high input tolerant and interference adaptive GaN LNA front-end, to an instinctual adaptive receiver. Fig. \ref{fig:concept} modifies the traditional RF receiver with the instinctual interference adaptation by incorporating 1) sensing through the onboard envelope detectors as observation points at the input and/or output of the GaN LNA, 2) processing through localized digital processing unit or the auxiliary ADC that's already present in baseband(BB), and 3) feedback to the customized GaN LNA bias controls for the best linearity and power performance while maintaining signal integrity. Note that in this paper, the processing is done using an MCU, but it can also be done using the BB DSP and auxiliary ADC. When blockers are present in a traditional Rx system working in nominal conditions, the blockers would saturate the LNA and produce a comparable IM3 to the actual signal which results in an undecidable LNA output. When blockers are present in the front-end with instinctual interference adaptation, the control logic would be able to increase the linearity of the system which in turn increases the IM3 compression at the cost of power consumption. When the signal and interference are both low, the Rx with bias control would be consuming less power for approximately the same signal levels. Note that the need for a better linearity range is present regardless of whether the high power is from the desired signal or the interference as long as the system is able to be brought back to the linearity range. 

% VIZOR did not use local feedback
% Orthogonal did not build a full system in hardware

% Columbia is extremely tightly integrated into the LNA, so the logic can't be too intelligent (like what one can do if a small processing unit beside the LNA makes decisions)

% Need for Instintual... DARPA Call... and current progress 

% Leading to our present work

% Many of the works have progressed toward a low-power adaptive receiver system. Some of works focused only on the circuit level ultra-low-power LNA design \cite{lowpowerLNA1,lowpowerLNA2} or an orthogonally tunable LNA design \cite{tunableLNA1} without a complete hardware system. Further, the circuit level LNA designs are more CMOS based whereas sometimes GaN are preferred in the industry. The dynamic bias illustrates that SNDR can also be tuned but again without the complete hardware system \cite{SNDR}. Additional works regarding an adaptive receiver system disregarded the speed of the tuning \cite{rxsystem} as well as a local feedback for fine tuning rather than just a LUT \cite{vizor,provizor}. 

To have a higher power handling capability and linearity, GaN LNA is utilized. Our prior work \cite{imspaper} involves interference adaptation for an Rx system with incremental adaptation control involving both feedforward and feedback path for the GaN LNA. However, the utilization of bench-top equipment significantly increases the adaptation time and the form factor of the system, which makes the design unsuitable in real life. \cite{imspaper} also lacks the consideration of the effects of the GaN LNA properties.  

This paper builds upon the prior work and has the following additional contributions:  
\begin{itemize}
    \item This work implements \textbf{the first sub-1ms interference adaptive, instinctual GaN LNA system} with a localized in-built intelligence using a microcontroller. The system consumes $\leq$10$\%$ of nominal LNA power to provide a wide tuning range of linearity for about 11 dB and LNA power for 0.5-2 W.
    \item The control circuitry of GaN LNA has been designed with careful consideration of the high-power effects of GaN LNA and the trade-off between system adaptation time and device lifetime.
    \item Background control theory of the system is provided on the limitations for the overall adaptation time (< 1 ms) which shows a high correlation with the measurement results. To the authors' current knowledge, this is the first control theory introduced for an interference adaptive RF front-end system.
    \item Important trade-offs are presented between two designs: 1) feedforward + feedback using control theory mentioned in Sec. \ref{sec:controltheory} and 2) feedback only using i) incremental adaptation, ii) lookup table (LUT) and iii) one-shot + incremental adaptations. The different designs illustrate different timing constraints and complexities for different applications.

\end{itemize}

% The paper further shows the speed of the adaptation is improved to the 1 ms range.   

The paper is organized as follows: Sec. \ref{sec:hardware} provides an overview of the control loops and component characterization;  Sec. \ref{sec:consideration} investigates different design considerations such as the design of directional coupler, high input power effects for a GaN LNA, Gate Voltage (V\textsubscript{G}) based tuning method and design comparison; Sec. \ref{sec:flow} describes the three control methods and the control theory; Sec. \ref{sec:results} presents the measurement results and Sec. \ref{sec:future} presents the future directions of this work.

% The very first letter is a 2 line initial drop letter followed
% by the rest of the first word in caps.
% 
% form to use if the first word consists of a single letter:
% \IEEEPARstart{A}{demo} file is ....
% 
% form to use if you need the single drop letter followed by
% normal text (unknown if ever used by the IEEE):
% \IEEEPARstart{A}{}demo file is ....
% 
% Some journals put the first two words in caps:
% \IEEEPARstart{T}{his demo} file is ....
% 
% Here we have the typical use of a "T" for an initial drop letter
% and "HIS" in caps to complete the first word.

%%%%%%%%%%%%%%%%%%%% \IEEEPARstart{I}{ntroduction} 

% You must have at least 2 lines in the paragraph with the drop letter
% (should never be an issue)

% \hfill mds
 
% \hfill August 26, 2015

\section{Hardware Description and Characterization}
\label{sec:hardware}
\subsection{System Architecture and PCB Designs}

\begin{figure}
\centering
\includegraphics[width=85mm]{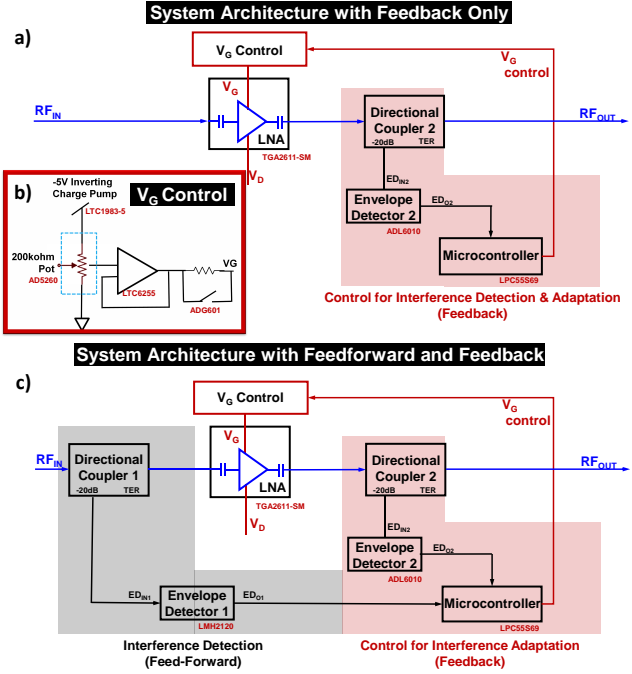}
\caption{a) System architecture with feedback only controls;  b) V\textsubscript{G} control circuitry; c) System architecture with feedforward and feedback controls.}
\vspace{-1.5em}
\label{fig:systemarch}
\end{figure}

\begin{figure}
\centering
\includegraphics[width=85mm]{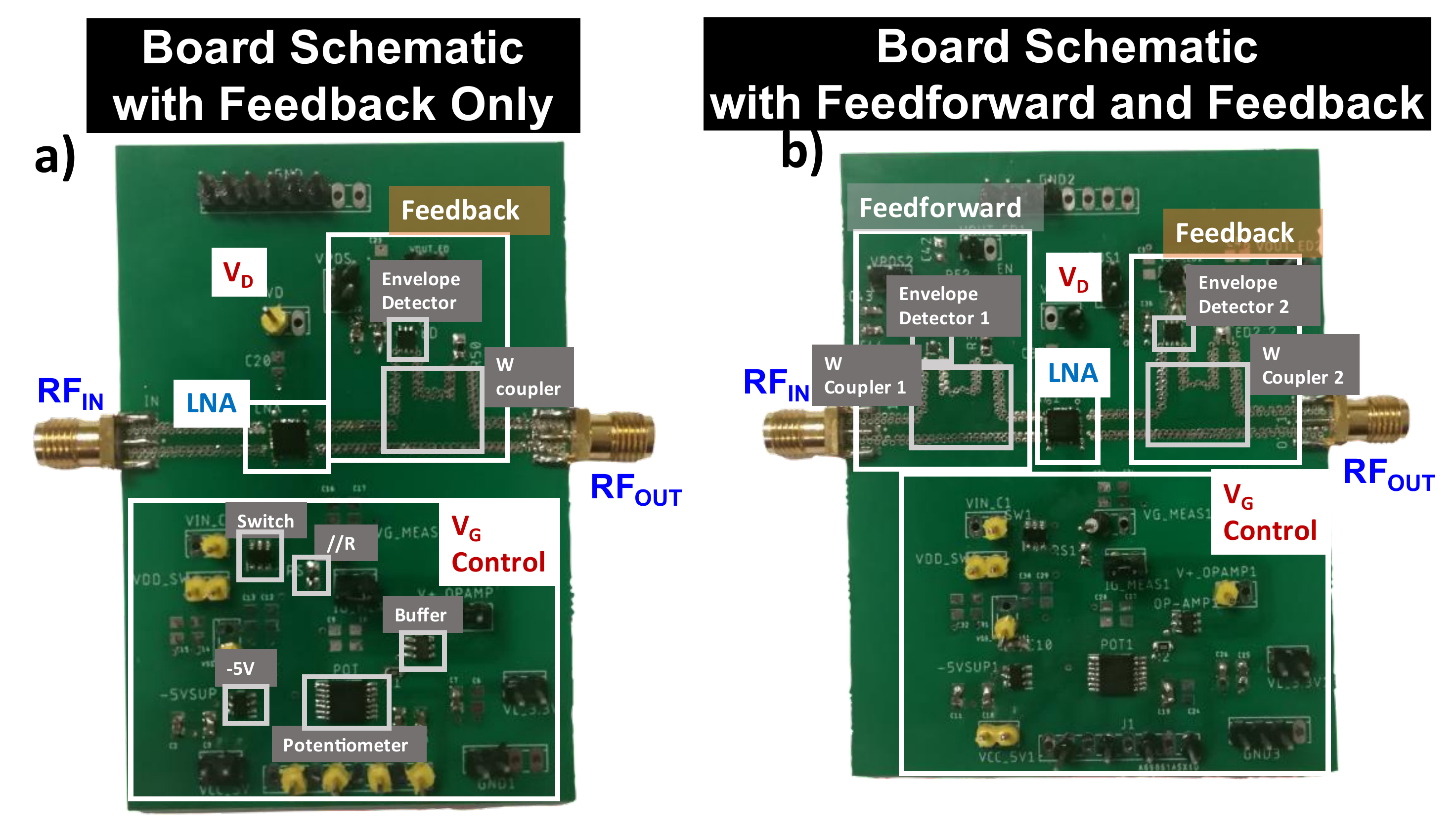}
\caption{a) Board schematic with feedback only controls; b) Board schematic with feedforward and feedback controls.}
\vspace{-1.5em}
\label{fig:boardsche}
\end{figure}

\begin{figure*}
\centering
\includegraphics[width=170mm]{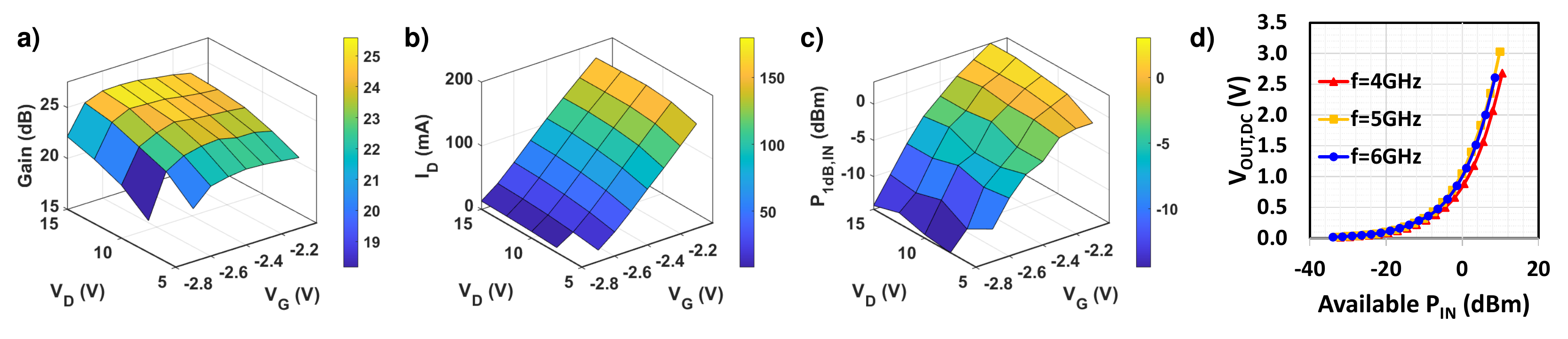}
\caption{Measured performance characterization of the components used: a) LNA gain vs. V\textsubscript{D} and V\textsubscript{G} at 4 GHz (weak function of both V\textsubscript{D} and V\textsubscript{G}), b) LNA drain current (I\textsubscript{D}) vs. V\textsubscript{D} and V\textsubscript{G} at 4 GHz (weak function of both V\textsubscript{D} and strong function of V\textsubscript{G}), c) LNA Input P\textsubscript{1dB} (P\textsubscript{1dB,IN}) vs. V\textsubscript{D} and V\textsubscript{G} at 4 GHz (weak function of both V\textsubscript{D} and strong function of V\textsubscript{G}), d) DC output voltage of the Envelope Detector 2 (ED2) at the output of the LNA vs. available input power to the ED.}
\vspace{-1.5em}
\label{fig:chara}
\end{figure*}

\begin{figure*}
\centering
\includegraphics[width=170mm]{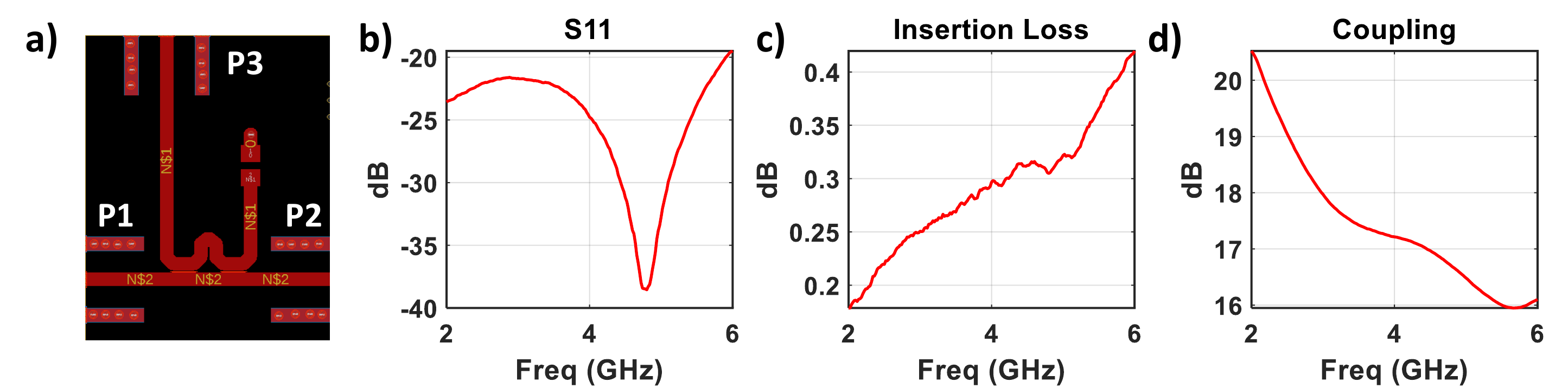}
\caption{W-shaped coupler a) schematic, b) S11, return loss, c) S21, insertion loss, d) coupling. }
% \vspace{-1.5em}
\label{fig:couplermeasure}
\end{figure*}

The two-layer PCB utilizes Rogers 4003C material with a thickness of 0.508 mm. The signal lines are carefully matched to 50 $\Omega$. The system architectures are shown in Fig. \ref{fig:systemarch} with associated board layouts shown in Fig. \ref{fig:boardsche}. The commercial off-the-shelf parts (COTS) used on the PCBs are listed in Table 1. Fig. \ref{fig:systemarch}a) and \ref{fig:boardsche}a) show the architecture and layout for the feedback-only design where the interference detection and adaptation are performed in the feedback loop that utilizes an envelope detector (ED) to sample the output of the LNA. The bias control circuitry for the LNA is shown in Fig. \ref{fig:systemarch}b). Fig. \ref{fig:systemarch}c) and \ref{fig:boardsche}b) show the architecture and layout for the feedforward and feedback design. The feedforward loop involves the use of envelope detector 1 (ED1) that's higher in sensitivity than envelope detector 2 (ED2) used in the feedback loop while still able to sustain the maximum input signal limited by the rating of the LNA at 30 dBm (1W). The difference between the two designs is how the interference will be detected (through the feedforward and feedback path, or directly from the feedback path) which will be discussed more in later sections.

The directional couplers and the LNA forms the front end in the system that will be connected to a mixer and baseband stages in a standard RF receiver. For a combined feedforward and feedback system, RF input first passes the through port of the directional coupler 1 to the LNA while the coupling port enables ED1 to measure the input signal level. Directional couplers are necessary for measurement because of the physical constraints (power handling capability) of the EDs as well as to avoid a significant power divide from the LNA. If the input signal is high (i.e. a blocker), then ED1 will output a higher DC voltage. The assumption is that high-level input signals are only caused by the interference and not the desired signal, thus high input signal corresponds to a high interference level. ED2 in the feedback path will also be incorporated in the interference detection for a better sensitivity after the LNA gain. The measurement will be processed through the ADC on the microcontroller and if the microcontroller determines the adaptation of the LNA is needed, tuning control will be initiated. The feedback control adjusts the gate voltage (V\textsubscript{G}) of the LNA to achieve the high linearity for the LNA which will be further discussed in Sec. \ref{sec:flow}. Note that the drain voltage (V\textsubscript{D}) of the LNA is not being controlled because the linearity improvement is minimal with changing V\textsubscript{D} for the specific LNA presented; however, V\textsubscript{D} controllability can be considered for future improvements on other LNAs. The difference with a feedback-only system is that the presence of the interference will only be detected using ED2 in the feedback loop.

\subsection{Characterization of Components} 

{% <-- We enclose the table in a group so that any redefinitions
%% are automatically undone at the end of the group.
%
\setlength{\tabcolsep}{1mm}%
\renewcommand{\arraystretch}{1.2}% for the vertical padding of table cells
\newcommand{\CPcolumnonewidth}{18mm}%
\newcommand{\CPcolumntwowidth}{19mm}%
\newcommand{\CPcolumnthreewidth}{45mm}%
\begin{table}[h]
\caption{Component Specifications}
\small% IMS: need this to get the 9pt text size in table cells % TODO is correct ?
\centering
\begin{tabular}{|l|l|l|}\hline
%\multirow{2}{6mm}{\parbox{8mm}{{\bfseries Font Size}}} & \multicolumn{3}{c|}{\raisebox{-0.25mm}{\bfseries Appearance (in Times New Roman or Times)}}\\ \cline{2-4}
\raisebox{-0.25mm}{\bfseries Component} & \raisebox{-0.25mm}{\bfseries Part Number} & \raisebox{-0.25mm}{\bfseries Specifications} \\ \hline
\parbox[t]{\CPcolumnonewidth}{\strut GaN LNA \strut}& \parbox[t]{\CPcolumntwowidth}{\strut TGA2611-SM (Qorvo) \cite{LNA}\strut} & \parbox[t]{\CPcolumnthreewidth}{\strut 2-6GHz, 1dB NF, 22dB Gain, -4dBm P\textsubscript{1dB,IN}, 1W nominal power\strut}\\ \hline
% \parbox[t]{\CPcolumnonewidth}{\strutDirectional\\Coupler\strut}& \parbox[t]{\CPcolumntwowidth}{MBDC\\-20-63HP \cite{DC}} & \parbox[t]{\CPcolumnthreewidth}{\strut2-6GHz, 0dB (through), -20dB (coupled), 100W max, 50$\Omega$\strut}\\ \hline
\parbox[t]{\CPcolumnonewidth}{\strut Envelope\\Detector 1\strut}& \parbox[t]{\CPcolumntwowidth}{LMH2120 \cite{ED1}} &\parbox[t]{\CPcolumnthreewidth}{\strut 0.05-6GHz, 7$\mu$s t\textsubscript{rise}, 2.9mA 0.032V-1.1V V\textsubscript{out}, 50$\omega$ P\textsubscript{IN}=-40$\sim$12dBm\strut}\\ \hline
\parbox[t]{\CPcolumnonewidth}{\strut Envelope\\Detector 2\strut}& \parbox[t]{\CPcolumntwowidth}{ADL6010 \cite{ED}} &\parbox[t]{\CPcolumnthreewidth}{\strut 0.5-43.5GHz, 47$\mu$s t\textsubscript{rise}, 1.6mA, 0.01V-3V V\textsubscript{out}, 50$\Omega$ P\textsubscript{IN}=-30$\sim$15dBm\strut}\\ \hline
\parbox[t]{\CPcolumnonewidth}{\strut -5V\\ Inverting Charge Pump\strut}& \parbox[t]{\CPcolumntwowidth}{LTC1983-5\cite{ICP}} &\parbox[t]{\CPcolumnthreewidth}{\strut -5V V\textsubscript{out}, 25$\mu$A \strut}\\ \hline
\parbox[t]{\CPcolumnonewidth}{\strut Digitally Programmed Potentiometer\strut}& \parbox[t]{\CPcolumntwowidth}{AD5260\cite{POT}} &\parbox[t]{\CPcolumnthreewidth}{\strut 200k$\Omega$, dual-supply, 256taps, 4-wire SPI, 0.3mW, t\textsubscript{settling}=5$\mu$s\strut}\\ \hline
\parbox[t]{\CPcolumnonewidth}{\strut Operational Amplifier\strut}& \parbox[t]{\CPcolumntwowidth}{LTC6255\cite{opamp}} &\parbox[t]{\CPcolumnthreewidth}{\strut 60$\mu$A, 6MHz GBP, 1.5V/us, 2.5$\mu$Vpp e\textsubscript{ni}\strut}\\ \hline
\parbox[t]{\CPcolumnonewidth}{\strut Switch\strut}& \parbox[t]{\CPcolumntwowidth}{ADG601\cite{sw}} &\parbox[t]{\CPcolumnthreewidth}{\strut 2.5ohm, 1$\mu$A, 80ns t\textsubscript{on}, 45ns t\textsubscript{off}, N.O., -60dB off isolation\strut}\\ \hline
\parbox[t]{\CPcolumnonewidth}{\strut Micro-\\controller\strut}& \parbox[t]{\CPcolumntwowidth}{LPC55S69\cite{micro}} &\parbox[t]{\CPcolumnthreewidth}{\strut 16bit ADC, 1MHz f\textsubscript{sample (ADC)}, SPI support, 150MHz f\textsubscript{CLK}\strut}\\ \hline
%12 & \parbox[t]{\CPcolumntwowidth}{{author name},\\author affiliation,\\email address\strut} & \\ \hline
%18 & title & \\ \hline
\end{tabular}
% \vspace{-1em}
\label{tab:COTS}
\end{table}
}% end of group enclosing the table

The components used are described in Table \ref{tab:COTS}. The LNAs in both the MWCL \cite{imspaper} and this paper are the same part number, but there are some chip-to-chip variations such that with the same V\textsubscript{G}, I\textsubscript{D} is higher in this paper. Note that the characteristics of the GaN LNA are determined more by the current rather than the bias voltage. This is because diodes are important in the GaN LNA modeling, and biasing the diode current with the correct voltage is more important \cite{LNAmodel}. We decided to continue with the V\textsubscript{G} range of -2.7 V to -2.2 V with a higher I\textsubscript{D} due to a worse S11 response at V\textsubscript{G} < -2.7 V as shown in Sec.\ref{sec:syscomparison}. The feedforward and feedback path components are characterized in the 2-6 GHz range. The losses due to the SMA cables are calibrated during the characterization of the RF components and measurements thereafter. RF components are chosen to have an input impedance of 50 $\Omega$. Fig. \ref{fig:chara}a)-c) represent the behavior of a GaN LNA at 4 GHz. As LNA's V\textsubscript{G} increases from -2.8 V to -2.2 V, the gain first increases and saturates at around -2.5 V, then starts to decrease. The gain also increases with increasing V\textsubscript{D}. Because the change in gain is less than 4 dB (neglecting the -2.8 V V\textsubscript{G} data due to low gain and P\textsubscript{1dB,IN}, the adaptation will be between V\textsubscript{G}=-2.7 V to -2.1 V), the LNA gain is a weak function of both V\textsubscript{D} and V\textsubscript{G}. The supply current (I\textsubscript{D}) increases by more than a hundred mA with increasing V\textsubscript{G} while having a small change with increasing V\textsubscript{D}. This makes LNA's I\textsubscript{D} a strong function of V\textsubscript{G} and a weak function of V\textsubscript{D}. The input P\textsubscript{1dB} (P\textsubscript{1dB,IN}) of the LNA varies around 16.5 dBm with increasing V\textsubscript{G} but changes minimally with V\textsubscript{D}. This makes the LNA's P\textsubscript{1dB,IN} a strong function of V\textsubscript{G} and a weak function of V\textsubscript{D}. Therefore, only V\textsubscript{G} will be changed in the later sections to achieve linearity while V\textsubscript{D} is fixed at 10 V. Note that if the LNA's P\textsubscript{1dB,IN} response is both a strong function of V\textsubscript{G} and V\textsubscript{D}, or if the LNA's gain is a strong function of V\textsubscript{D}, both V\textsubscript{G} and V\textsubscript{D} can be implemented in the control loop.

Fig. \ref{fig:chara}d) represents the behavior of ED2 at frequencies of 4, 5, and 6 GHz. The voltage output of ED2 increases exponentially with increasing power. ED1 is chosen to be able to detect low power levels at the input. Other controlling components listed in Table \ref{tab:COTS} are generally chosen to be low power, short settling and rising time, and low noise.

% All directional couplers presented are identical. The couplers are necessary due to the physical constrains of the EDs. 
% The directional couplers are identical in design where the insertion loss is around ...0.5dBm..., the coupling to ED1 is around ...20dBm..., and the termination port is terminated with 50 $\Omega$.

\section{Design Considerations}
\label{sec:consideration}

\begin{figure}
\centering
\includegraphics[width=70mm]{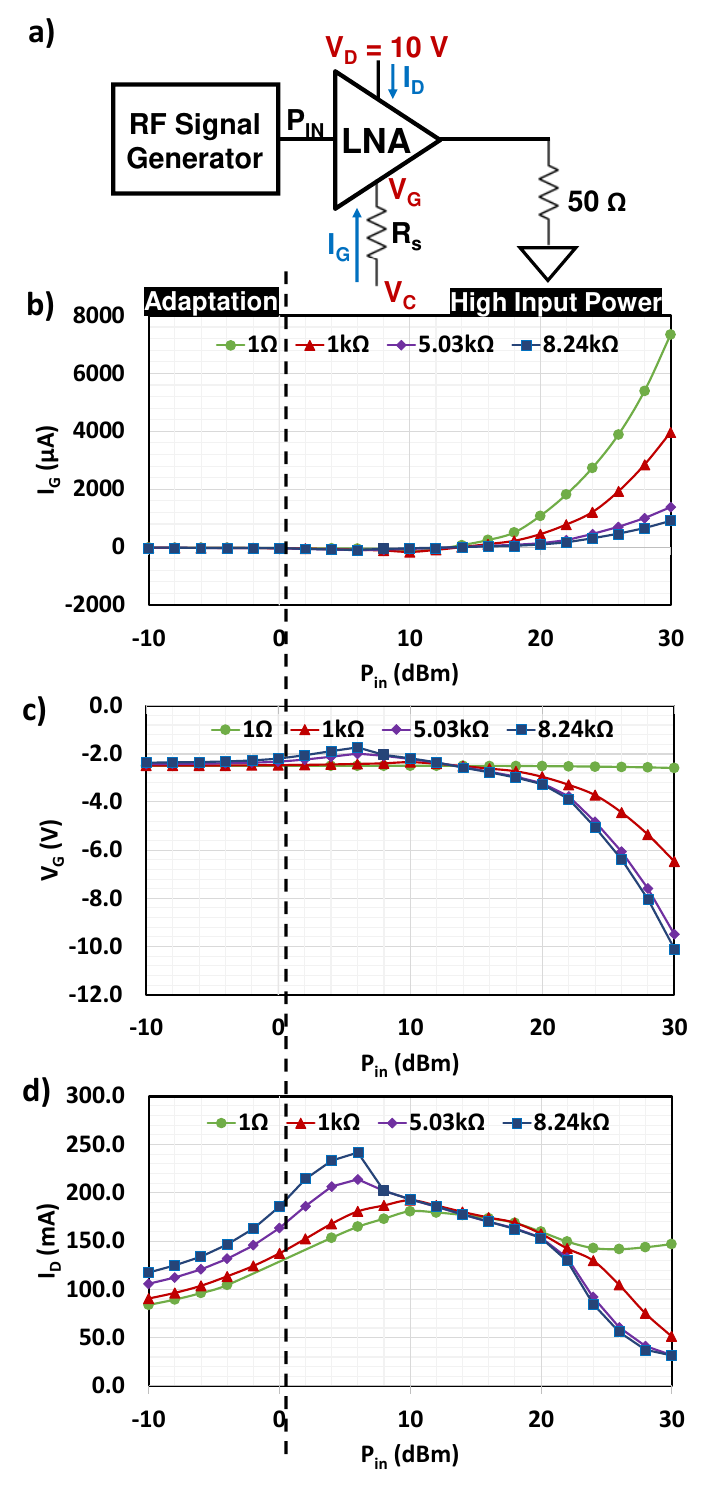}
\caption{2 GHz LNA at V\textsubscript{G}=-2.5 V and V\textsubscript{D}=10 V data with various series resistors at the bias gate node of the LNA for a) LNA high power measurement set up; b) I\textsubscript{G} vs. P\textsubscript{IN} adjusted for cable loss; c) V\textsubscript{G} vs. P\textsubscript{IN}; d) I\textsubscript{D} vs. P\textsubscript{IN}. }
\vspace{0em}
\label{fig:VGHighPower}
\end{figure}

\subsection{Directional Coupler Design}

% \begin{figure}
% \centering
% \includegraphics[width=75mm]{figures/couplers.pdf}
% \caption{a) U-shaped directional coupler; b) W-shaped directional coupler; c) double sided U-shaped directional coupler; d) double sided W-shaped directional coupler}
% \vspace{-1.5em}
% \label{fig:couplers}
% \end{figure}

% \subsubsection{Directional Coupler Considerations}

In our prior work \cite{imspaper}, we utilized COTS directional coupler components to provide the measurement path; however, even though the components have outstanding specs, the integration of the directional coupler with the PCB board is causing unwanted reflections from the soldering and the abrupt transition from the trace to the component. Consequently, this paper takes advantage of the onboard microstrip design for the directional coupler which avoids extra transition from the board to the component. The coupler is necessary for decreasing the power input to below the power limit of the envelope detector and not to diverge extra power from the signal path for measurement purposes. 
The microstrip directional coupler schematic is shown in Fig. \ref{fig:couplermeasure}a) \cite{couplerref}. The measurements for the directional coupler are shown in \ref{fig:couplermeasure}b)-d). The S11 shows low return loss with measurements below -19 dB over the frequency from 2 GHz to 6GHz. The insertion loss (S21) increases from 0.18 dB to 0.42 dB with frequency. The coupling (S31) changes from 20.5 dB to 16 dB over the bandwidth.

% The directivity of the couplers is an important consideration during the design process because as directivity decreases, forward power measurement errors can be as high as several dB, thus as a rule of thumb, the desired directivity should be above 15 dB. The directivity is measured using the formula D = S31 + S21 - S32 which translates to coupling + insersion - isolation\cite{directivity}.

% \subsection{Addition of Buffer and Switch//R Network}
% trade off between settling time and 

\begin{figure}
\centering
\includegraphics[width=80mm]{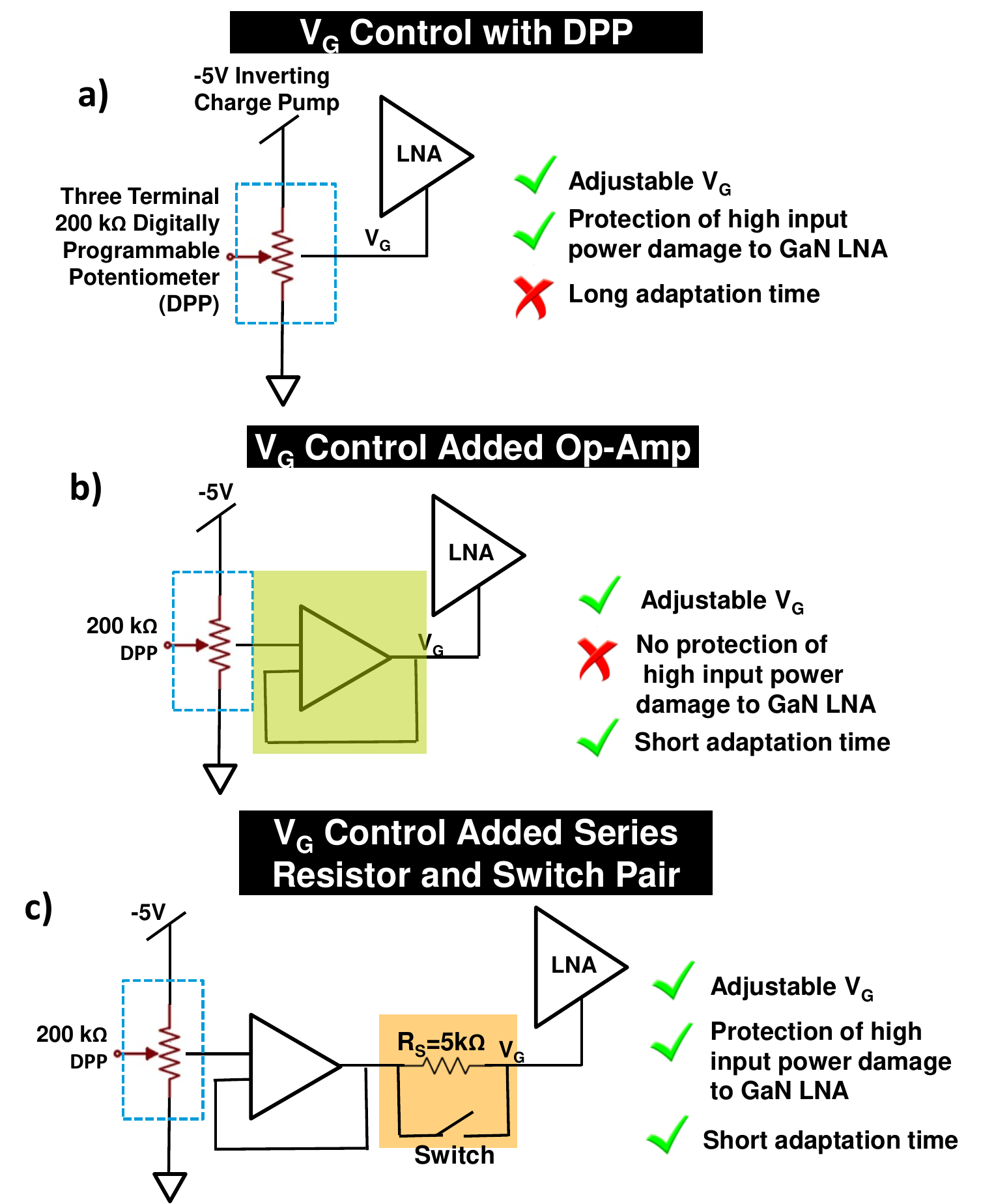}
\caption{a) V\textsubscript{G} controller with only the digitally programmable potentiometer; b) V\textsubscript{G} controller with a digitally programmable potentiometer with a buffer; c) current V\textsubscript{G} controller design with buffer and resistor and switch pair to accommodate both low and high input power to the LNA.}
% \vspace{-1.5em}
\label{fig:bfsw}
\end{figure}

\subsection{GaN LNA Considerations for High Power Input}
\label{sec:Hpower}

% (Maybe rewording??)

Unlike the typical GaAs LNA's lifetime being limited by the breakdown voltage, GaN LNA is limited by high DC gate current and drain-gate voltage (V\textsubscript{DG}). Characteristics of a GaN LNA with high input power have been investigated in \cite{LNAsurvive}. Similar characteristics have been presented for the LNA used in this paper in Fig. \ref{fig:VGHighPower}.

Fig. \ref{fig:VGHighPower} describes I\textsubscript{G}, V\textsubscript{G} and I\textsubscript{D} vs. different input power with different resistor values in series (R\textsubscript{S}) at the gate bias of the LNA with the set up in Fig. \ref{fig:VGHighPower}a). The frequency has been set at 2 GHz for the worst-case scenario. As shown in the plots \ref{fig:VGHighPower}b)-d), low power characteristics are very different than the high power characteristics. In the low power region or the adaptation range, I\textsubscript{G} is actually negative causing V\textsubscript{G} to tail up due to R\textsubscript{S}, and with Ohm's law, when I\textsubscript{G} is flowing out of the gate, V\textsubscript{G} is higher than the control voltage (V\textsubscript{C}). Because of the higher bias in V\textsubscript{G}, I\textsubscript{D} is also drawing more current. With increasing in R\textsubscript{S}, V\textsubscript{G} tails up higher and I\textsubscript{D} also reaches a higher point. When the input power transitions to the high power region, I\textsubscript{G} becomes positive and starts to increase exponentially. As a result of I\textsubscript{G} becoming more positive, V\textsubscript{G} starts to drop off exponentially with I\textsubscript{D} also dropping. With increasing in R\textsubscript{S}, I\textsubscript{G} increases more gradually which protects the LNA. Even though V\textsubscript{G} drops to about -10 V range, it is not significant to cause the LNA to break down. The typical critical value for V\textsubscript{DG} causing degradation in LNA performance is 30 V, where if V\textsubscript{D} is 10 V (used in this paper), the minimum V\textsubscript{G} is -20 V \cite{LNAsurvive3}. 

The characteristics of I\textsubscript{G} with varying P\textsubscript{IN} is explained by the Shockley contact at the gate and source of a GaN high-electron-mobility transistor (HEMT). A GaN LNA model can be found in \cite{LNAmodel}. The Shockley contact is essentially a diode, consequently, at low power levels, the gate experiences a small leakage current; at high power levels, the diode is turned on and I\textsubscript{G} increases exponentially \cite{LNAsurvive2}.  

Another effect of the GaN LNA is the trapping effect on the gain recovery after a pulse of high input signal \cite{EW}. This effect happens much higher than the adaptation range (P\textsubscript{IN} > 25 dBm), thus during the pulse, the system can be tuned at the maximum possible linearity. When the high input signal is removed, the slow gain recovery does not affect the detection that the input signal is now minimal and returns back to the low power mode.

\begin{figure}
\centering
\includegraphics[width=85mm]{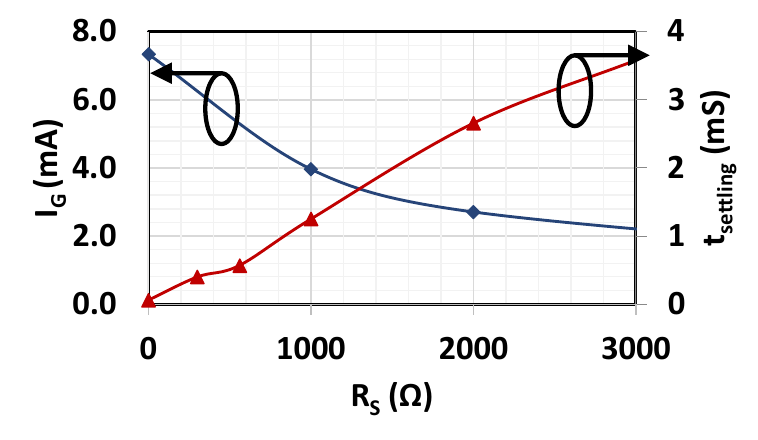}
\caption{2GHz data at V\textsubscript{G}=-2.5 V and V\textsubscript{D}=10 V for the trade-off of I\textsubscript{G} and settling time vs. different series resistor value.}
\vspace{-1.5em}
\label{fig:Rseriessettling}
\end{figure}

\subsection{Bias Control }

% P\textsubscript{IN}-->LNA-->EDin-->VEDOUT-->Potentiometer Adjustment--> Buffer -->Switch//R-->VG adjustment
To protect the GaN LNA against the high input power characteristics and provide reasonable adaptation time, the current design for tuning the V\textsubscript{G} of the LNA is proposed in Fig. \ref{fig:bfsw}c). The microcontroller will be configuring the digitally programmable potentiometer (DPP) which is supplied by a -5 V inverting charge pump for the negative V\textsubscript{G} bias. Following the DPP is a buffer and a parallel structure of a resistor and switch for the control of I\textsubscript{G}.

The process for the current design in tuning the V\textsubscript{G} is shown in Fig. \ref{fig:bfsw}. Fig. \ref{fig:bfsw}a) considers the case when there is only a DPP connected directly to the gate of the LNA. The drawback of this design is shown in Fig. \ref{fig:Rseriessettling}. As the R\textsubscript{S} to the gate of the LNA increases, the gate current is suppressed but the settling time increases. Since the DPP is in the 100 k$\Omega$ range, the high series resistance causes a high RC time constant which increases the settling time (T\textsubscript{S}) significantly. The increase in Ts would, in turn, increase the adaptation time to tens of milliseconds which is undesirable when one of the goals is to constraint the timing in the millisecond range. To solve the Ts problem, a buffer is added  as shown in Fig. \ref{fig:bfsw}b) which essentially reduces R\textsubscript{S} to decrease T\textsubscript{S}; however, as discussed in Sec. \ref{sec:Hpower}, the reduced R\textsubscript{S} draws more I\textsubscript{G} in high power which minimizes the device lifetime.

To balance the I\textsubscript{G} and Ts trade-off, a parallel resistor-switch pair is added in the current design as shown in Fig. \ref{fig:bfsw}c). During the low power and adaptation regions in Fig. \ref{fig:VGHighPower}a), I\textsubscript{G} is low in the sub-mA region, so the switch can be closed to construct the low resistance path for tuning V\textsubscript{G} to improve Ts. After the adaptation, the switch will remain closed to maintain a steady bias condition, on the other hand, if the switch opens to place R\textsubscript{S} in the path, V\textsubscript{G} would increase as shown in Fig. \ref{fig:bfsw}b) and change the bias condition which consumes more power. In the high power region, lower I\textsubscript{G} is more prominent since V\textsubscript{G} is set to -2.1 V directly without the adaptation, so T\textsubscript{S} can be traded for lower I\textsubscript{G} with the switch being open and resistor in series.
% ....MAYBE KEEP CLOSE FOR CHANGE TO -2.1 THEN SWITCH OUT....

\subsection{Linearity Decision Making in Incremental Adaptation}

\begin{figure}[h]
\centering
\includegraphics[width=70mm]{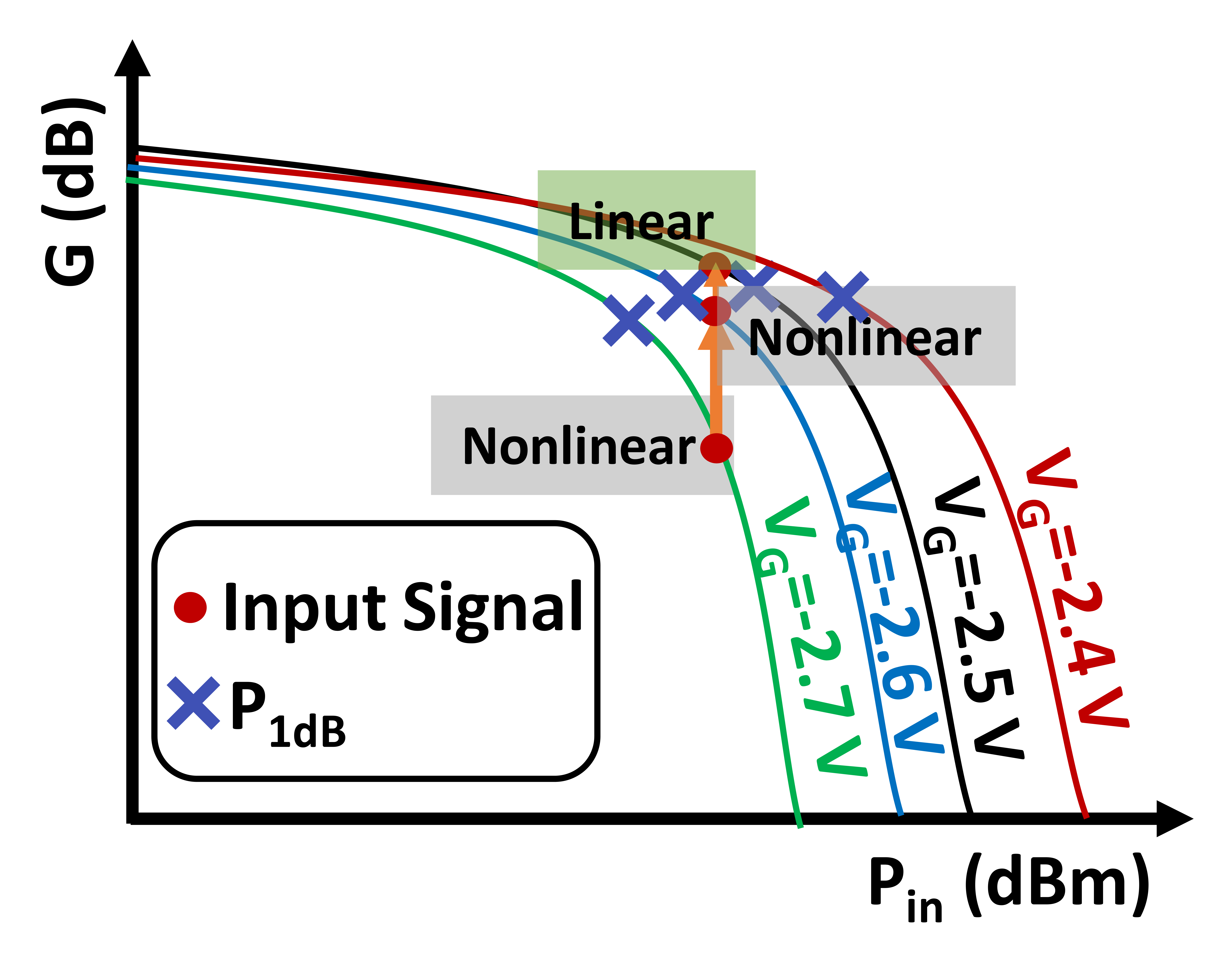}
\caption{Conceptual diagram for the gain of the system vs. P\textsubscript{IN}. The input signal moves from nonlinear region to linear region of the system by increasing V\textsubscript{G} from -2.7 V to -2.5 V.}
\vspace{0 em}
\label{fig:tuningex}
\end{figure}

Due to the variable gain behavior of the LNA with changing V\textsubscript{G}, the ED2 measurements at the output of the LNA are also affected. The gain of the LNA first increases with increasing V\textsubscript{G}, then decreases after V\textsubscript{G}$\approx$-2.5 V in Fig. \ref{fig:chara}a). Fig. \ref{fig:tuningex} shows that the linear region occurs where the gain of the system remains relatively constant before the 1dB compression point (P\textsubscript{1dB}) and the nonlinear region occurs where the system is highly compressed beyond P\textsubscript{1dB}. Even though the gain varies with increasing V\textsubscript{G}, P\textsubscript{1dB} monotonically increases with increasing V\textsubscript{G}. When a high input signal presents in the system at V\textsubscript{G}=-2.7 V, the signal is highly gain compressed and the system is highly nonlinear. To linearize the system, V\textsubscript{G} increases in 0.1 V increments. When V\textsubscript{G} increases from -2.7 V to -2.6 V, the signal becomes less gain compressed but the system remains nonlinear. To further improve the system linearity, V\textsubscript{G} is increased to -2.5 V. The idea of incrementing V\textsubscript{G} to improve linearity forms the basis of the incremental adaptation in Sec. \ref{sec:incadapt}. 

\begin{figure}
\centering
\includegraphics[width=80mm]{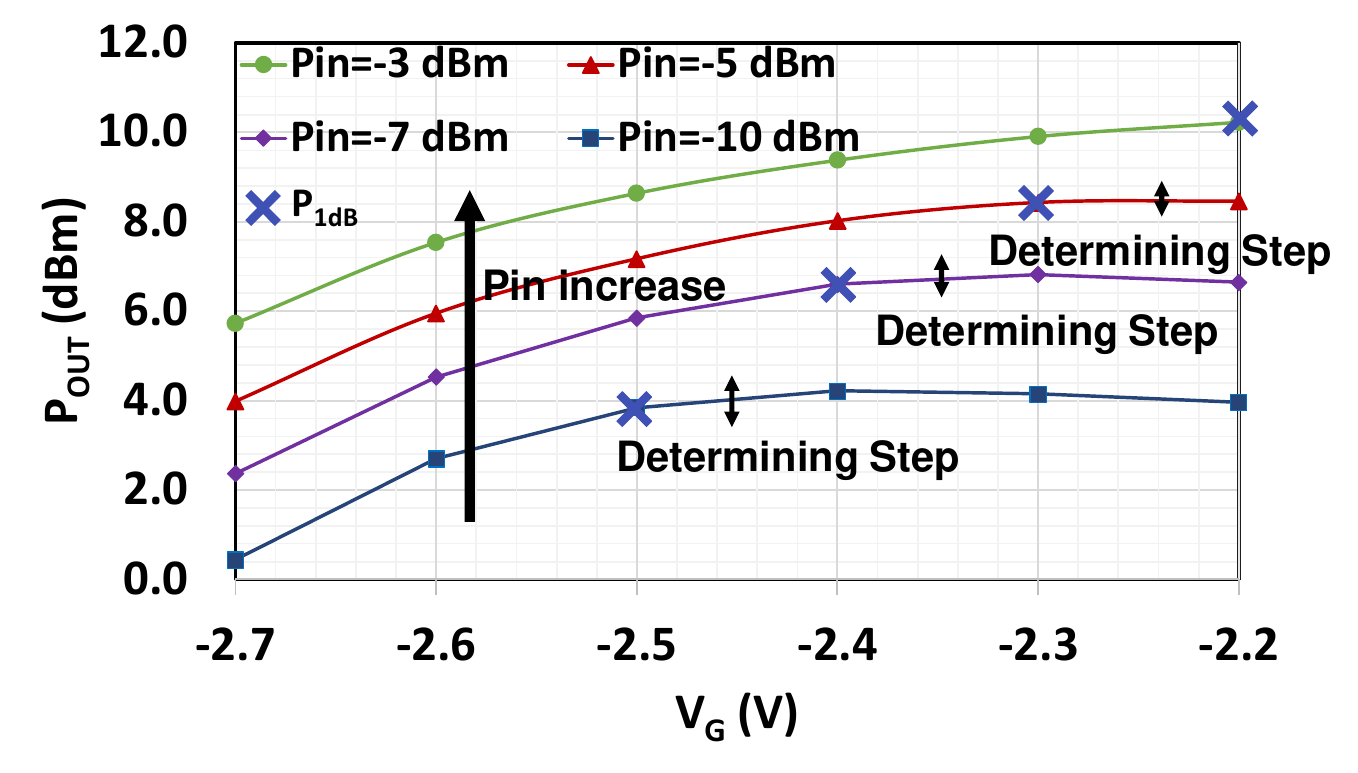}
\caption{3 GHz data for the output power of the LNA vs. V\textsubscript{G} at different input power levels and the P\textsubscript{1dB} points signifying the expected V\textsubscript{G} to bring the system back to linearity for each input power. The determining steps show how the threshold applied in the control logic is decided.}
\vspace{-1.5em}
\label{fig:DecisionVG}
\end{figure}

Fig. \ref{fig:DecisionVG} shows P\textsubscript{OUT} vs V\textsubscript{G} tuning with respect to different P\textsubscript{IN}. P\textsubscript{1dB}s are also shown to show that V\textsubscript{G} < V\textsubscript{P\textsubscript{1dB}} results in a nonlinear system. As P\textsubscript{IN} increases, the V\textsubscript{P\textsubscript{1dB}} also increases. As V\textsubscript{G} is tuning, the gain of the LNA increases then decreases as shown in Fig. 4a). Due to the effect of LNA gain with different VG and LNA transitioning from non-linear to linear, the determining step decreases with P\textsubscript{IN} increase. Therefore to accommodate the difference in determining steps in both high and low P\textsubscript{IN} across different frequencies, a lower determining step is chosen as the linearity threshold which has a drawback of overestimation of V\textsubscript{G} for lower P\textsubscript{IN} values. One may suggest that different threshold limits can be used with different P\textsubscript{IN} values, but this again creates a case-dependent threshold that may not work in other frequencies.

\section{Control Mechanisms for Adaptation}
\label{sec:flow}

Fig. \ref{fig:incrementaladapt}-\ref{fig:lutincrementaladapt} present three control methods in the feedback only configuration. 

\subsection{Incremental Adaptation Control Logic}
\label{sec:incadapt}

\begin{figure*}
\centering
\includegraphics[width=180mm]{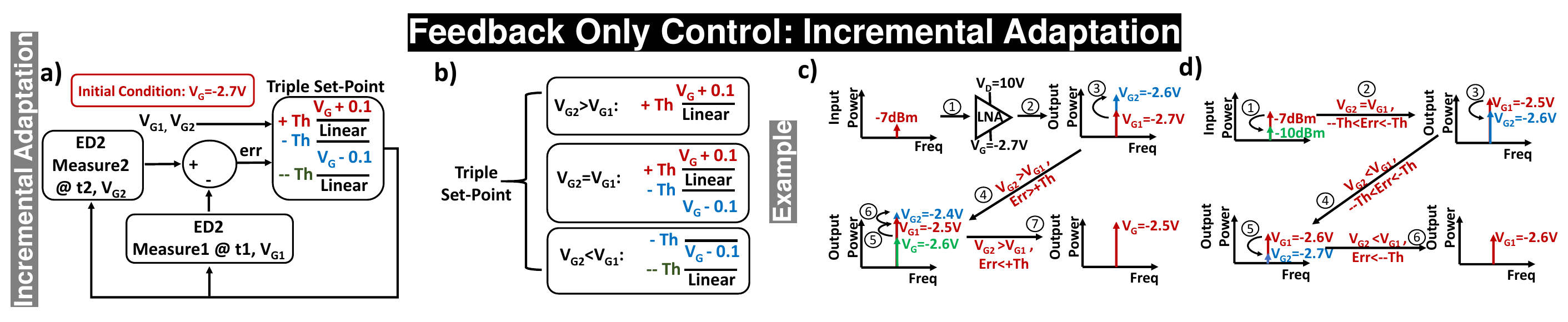}
\caption{a) Incremental adaptation logic for feedback only control; b) triple set-point breakdown for different V\textsubscript{G} cases; c) example of the incremental adaptation when a detectable interference presents; d) example of the incremental adaptation when interference level decreases.}
\vspace{-1.5em}
\label{fig:incrementaladapt}
\end{figure*}

\begin{figure*}
\centering
\includegraphics[width=180mm]{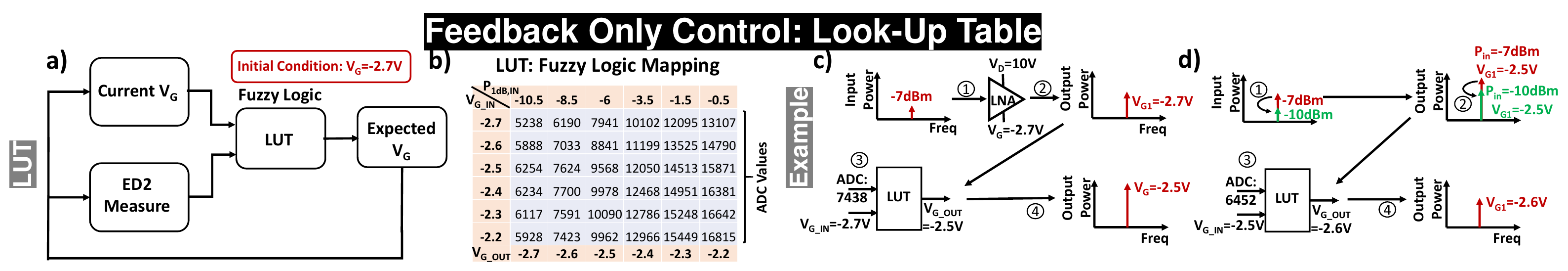}
\caption{a) Look-up table (LUT) logic for feedback-only control; b) LUT mapping for different V\textsubscript{G} and input powers; c) example of the LUT adaptation when a detectable interference presents; d) example of the LUT adaptation when interference level decreases.}
\vspace{-1.5em}
\label{fig:lutadapt}
\end{figure*}

\begin{figure*}
\centering
\includegraphics[width=180mm]{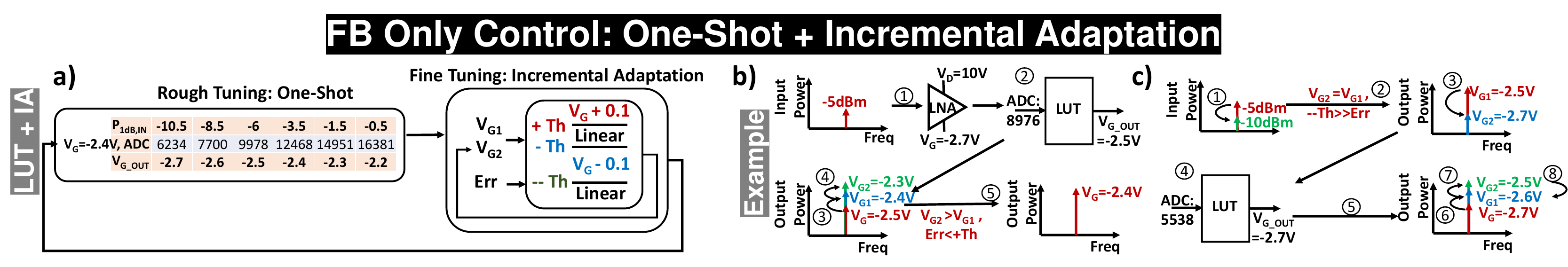}
\caption{a) One-shot + incremental adaptation logic for feedback only control; b) example of the one-shot + incremental adaptation when a detectable interference presents; c) example of the one-shot + incremental adaptation when interference level decreases.}
\vspace{-1.5em}
\label{fig:lutincrementaladapt}
\end{figure*}

Fig. \ref{fig:incrementaladapt}a) describes the control method with incremental adaptation. The LNA is initially in low power mode with V\textsubscript{G} being initially set at -2.7 V and ED2 will perform the initial measurement at time instance 1 (t1) as the reference. When interference reaches the system, another ED2 measurement at time instance 2 (t2) is taken as the current value. The difference (err) between the current and reference of ED2 is compared with a set of thresholds to start the feedback control. The set of thresholds can be regarded as an extended version of bang-bang control as the triple set-point control in Fig. \ref{fig:incrementaladapt}b). The triple set-point has three different conditions: when V\textsubscript{G1} at t1 is the same as V\textsubscript{G2} at t2, the resulting action is increment V\textsubscript{G} by 0.1 V if the err is greater than the positive threshold (+Th), decrement V\textsubscript{G} by 0.1 V if the err is less than the negative threshold (-Th), or maintain the same V\textsubscript{G} if the err is between +Th and -Th; when V\textsubscript{G2} is greater than V\textsubscript{G1} (VG is incrementing), the err is compared only to the +Th such that V\textsubscript{G} incrementes when err is greater than +Th (gain compression observed) or returns to V\textsubscript{G1} when err is less than +Th (linear); when V\textsubscript{G2} is less than V\textsubscript{G1} (VG is decrementing), the err is compared with the -Th and a more negative threshold (--Th) such that V\textsubscript{G} decrements to find the optimum V\textsubscript{G} when the err is in between the thresholds (still in the linear region), or returns to V\textsubscript{G1} when the err is less than the --Th (gain compression observed). An example of the incremental adaptation is shown in Fig. \ref{fig:incrementaladapt}c). When the interference is detected, V\textsubscript{G} will start incrementing until err is less than the +Th. Another example of when the interference signal changes are shown in \ref{fig:incrementaladapt}d). When the interference level decreases, the V\textsubscript{G} also starts to decrement until err is less than --Th. Note that if the interference level drops a significant amount, V\textsubscript{G} is back to the initial condition instead of stepping down to start the adaptation. 

\subsection{Look-Up Table Control Logic}
Fig. \ref{fig:lutadapt}a) describes the control method with a look-up table (LUT). A LUT as shown in Fig. \ref{fig:lutadapt}b) utilizes the current V\textsubscript{G} value and the ADC measurement (unit: 10$^{-4}$ V) of the ED2 at 3 GHz, then outputs the V\textsubscript{G} value that would bring the LNA back to linearity (High linearity mode). Note that, although not included in the figure, after settling of V\textsubscript{G}, ED2 is constantly monitored so that if the measurement is within the ADC variation, V\textsubscript{G} stays at the same value; otherwise, the process starts over with another ED2 measurement. The LUT is generated in a way that for each input power using the P\textsubscript{1dB,IN} at each V\textsubscript{G} value and different V\textsubscript{G} settings, ADC measures at the output of ED2. LUT can be associated with a Fuzzy logic control that unlike the triple set-point control to only have four commands, Fuzzy logic includes a wider range of V\textsubscript{G} outputs. Each combination input V\textsubscript{G} and ED2 measurement can be treated as an if-else statement in the Fuzzy logic and returns a preprogrammed output V\textsubscript{G} \cite{fuzzy}. An example of the LUT control is shown in \ref{fig:lutadapt}c). When the interference is detected, ADC measures at ED2 output. With the ADC measurement of 743.8 mV (ADC reads 7438) and the current V\textsubscript{G} value of -2.7 V, the LUT determines that a V\textsubscript{G} value of -2.5 V with a P\textsubscript{1dB,IN} of -6 dBm is sufficient to bring the system back to linearity. Another example is shown in \ref{fig:lutadapt}d) in the case of reduced interference level. Again the ADC measurement and current V\textsubscript{G} value is used to determine that a V\textsubscript{G} of -2.6 V with a P\textsubscript{1dB,IN} of -8.5 dBm is sufficient for linearity.

\subsection{One-Shot + Incremental Adaptation Control Logic}
Fig. \ref{fig:lutincrementaladapt}a) describes the control method with a one-shot for rough tuning and incremental adaptation for fine-tuning. The one-shot is implemented in a LUT style with certain degrees of an underestimate of the output V\textsubscript{G} value to accommodate different frequencies and different V\textsubscript{G} values. The incremental adaption again utilizes the triple set-point control with three thresholds and four regions of action. An example is shown in \ref{fig:lutincrementaladapt}b) where when the interference of -5 dBm is detected, one-shot rough tunes the V\textsubscript{G} to -2.5 V, then the incremental adaptation starts to increment V\textsubscript{G} for fine-tuning. A second example is shown in \ref{fig:lutincrementaladapt}c) where when the interference level drops significantly, one-shot tunes V\textsubscript{G} to -2.7 V, and then incremental adaptation steps up to find the optimum V\textsubscript{G}.

% Also note here that unlike LUT where the ED2 is constantly monitored to compare with the ADC variation, more change in the ED measurement is required to start over the cycle. The increase in change of the ED measurement is necessary to avoid false triggering and instability in the loop due to the less accurate tuning of V\textsubscript{G}. 

\subsection{Comparison of Control Methods}
All three methods are intended to control the bias of LNA in order to achieve linearity with a minimum required power consumption from the LNA. LUT has the advantage of being very fast and accurate if the frequency is known so that the specific LUT can be utilized; however, accuracy implies a stringent requirement on the memory space in the microcontroller that multiple LUTs at different frequencies are required. On the other hand, the incremental adaptation only and the one-shot + incremental adaptation methods do not need the frequency information for adaptation as long as the measurement is greater than the interference threshold. However, due to the extra adaptation and the single set linearity threshold, the two methods tend to take longer and are less accurate involving some overestimation of V\textsubscript{G}. A way to incorporate methods is that when the interference frequency is known, LUT can be used, but when the frequency is unknown, one-shot + incremental adaptation can be used.

% P\textsubscript{OUT} act-exp or difference
% LNA: P\textsubscript{OUT} increases then decrease with increase in V\textsubscript{G}. Decision pout change decreases with increasing P\textsubscript{IN}.
% ED: Exponential Vout vs. P\textsubscript{IN}

\subsection{Control Theory for Optimum Bias}
\label{sec:controltheory}
\begin{figure}[h]
\centering
\includegraphics[width=85mm]{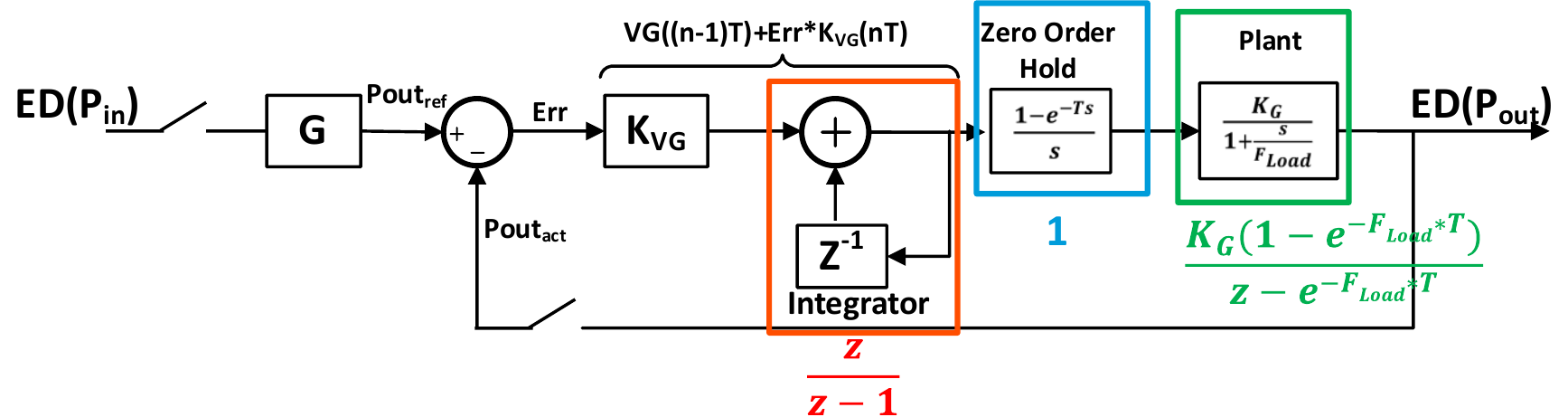}
\caption{Second order dynamic control diagram with discrete time control and continuous time plant.}
% \vspace{-1.5em}
\label{fig:controltheory}
\end{figure}

In order to provide a more thorough background on the limitations of the control loop adaptation time, a second-order dynamic system is presented in Fig. \ref{fig:controltheory} following analysis for digital dropout regulators \cite{controltheory1,controltheory2}. The control loop models the feedforward and feedback control method with two degrees of observability at the input and output. The model has the following assumptions:
\begin{enumerate}
  \item flat gain until the P\textsubscript{1dB} point.
  \item gain is not a function of V\textsubscript{G} and frequency but linearity.
  \item V\textsubscript{G} settling (100kHz) dominates ED bandwidth (40 MHz) and LNA bandwidth.
\end{enumerate}

The goal of the control loop is to minimize the difference between the expected output and the measured output. The expected output is calculated from the measured input and multiplied by a constant LNA gain (G). Both the expected output and the measured output are sampled and subtracted to form an error signal. The error signal multiplies with a proportional constant (K\textsubscript{VG}) to form a $\Delta$ V\textsubscript{G} that is added to the previous V\textsubscript{G} through the integrator (\( \frac{z}{z-1} \)). The sampled V\textsubscript{G} transforms to continuous time through the zero-order hold. Finally, V\textsubscript{G} supplies to the plant and maps the V\textsubscript{G} to the measured output. The plant consists of a constant gain K\textsubscript{G} and a pole from the V\textsubscript{G} settling time (F\textsubscript{Load}) with a z-domain equation of \( \frac{K\textsubscript{G}(1-e^{(-F_{Load}T_{s})})}{z-e^{(-F_{Load}T_{s})}} \). T\textsubscript{s} is the ADC sampling period(we will approximate to 50$\mu$s for simpler demonstration of calculation). The open loop gain is
\begin{equation}
   G_{OL}(z)=\frac{K_{VG}K_G(1-e^{-F_{Load}T_{s}})}{z-e^{(-F_{Load}T_{s})}}\times\frac{z}{z-1}.
\end{equation}

\begin{figure}
\centering
\includegraphics[width=85mm]{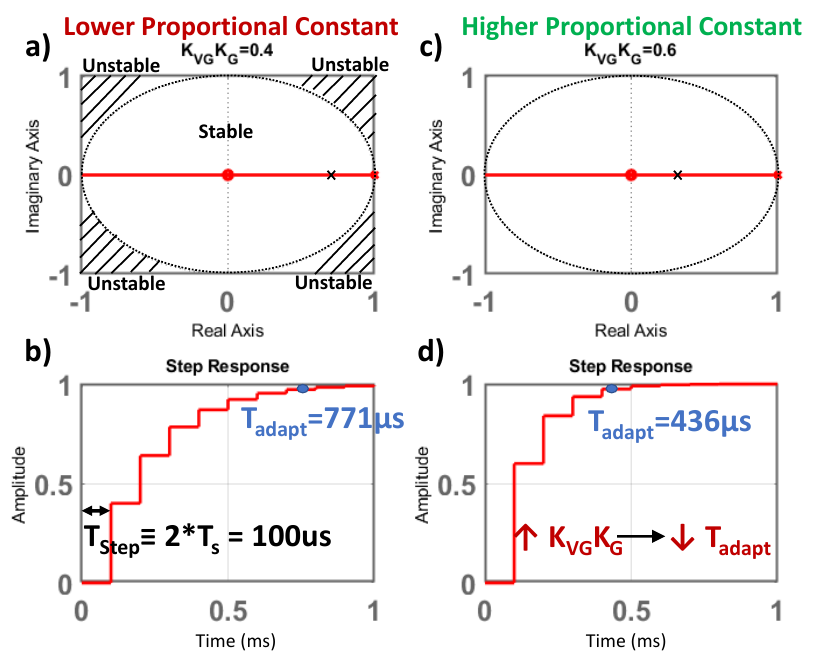}
\caption{MATLAB simulation (root locus and step response) showing the effect on the step response of the system with a different proportional gain constant and 2*T\textsubscript{s}=100 $\mu$s.}
% \vspace{-1.5em}
\label{fig:controltheoryplot}
\end{figure}

To maximally match the existing control methods, since each adaptation step requires two samples of measurement to ensure accuracy which will be explained shortly, 2*T\textsubscript{s}=100 $\mu$s with a sampling frequency F\textsubscript{s}/2=10 kHz is used. With F\textsubscript{Load}=100 kHz, F\textsubscript{s} <  F\textsubscript{Load}, F\textsubscript{s} dominants in the settling time. From Fig. \ref{fig:controltheoryplot}a)-b), the proportional constant is low with K\textsubscript{V\textsubscript{G}}K\textsubscript{G}=0.4, the total settling time or the adaptation time is 771 $\mu$s. From Fig. \ref{fig:controltheoryplot}c)-d), K\textsubscript{V\textsubscript{G}}K\textsubscript{G}=0.6, the total settling time is 436 $\mu$s. As shown in the step response, because T\textsubscript{Load} < T\textsubscript{s}, each step takes 2*T\textsubscript{s}. The different proportional constant can be thought of as the one-shot values in the one-shot + incremental adaptation method, with a higher one-shot value (high K\textsubscript{V\textsubscript{G}}K\textsubscript{G}), the number of steps to reach a steady state is lower, hence a faster response.

Overall, the adaptation time can be approximated as 
\begin{equation}
\label{eq:adapt}
   T_{adapt}= N * (2*T_{s} + T_{process}),
\end{equation}
where
\begin{equation}
\label{eq:Ts}
    T_{s} > T_{V_G} + T_{LNA} + T_{ED}.
\end{equation}
In the equations, N is the number of steps which is a function of the proportion constant, T\textsubscript{s} is the ADC sampling time, T\textsubscript{process} is the processing time for the microcontroller to make decisions, T\textsubscript{V\textsubscript{G}} is the V\textsubscript{G} tuning time, T\textsubscript{LNA} is the propagation time from tuned V\textsubscript{G} to the settling of the LNA and T\textsubscript{ED} is the propagation time from settled LNA output to settled ED output. Note that the first sample of the ADC contains some of the transient response which lacks accuracy, so the second sample is taken to be the accurate one to the processing, hence the 2*T\textsubscript{s} in Eq.\ref{eq:adapt}. In order for the second sample to be accurate, the first sample of the ADC should contain all of the transient responses, hence Eq.\ref{eq:Ts} shows that the sampling time of the ADC needs to be greater than the total propagation and settling time for each component.

\begin{figure}
\centering
\includegraphics[width=85mm]{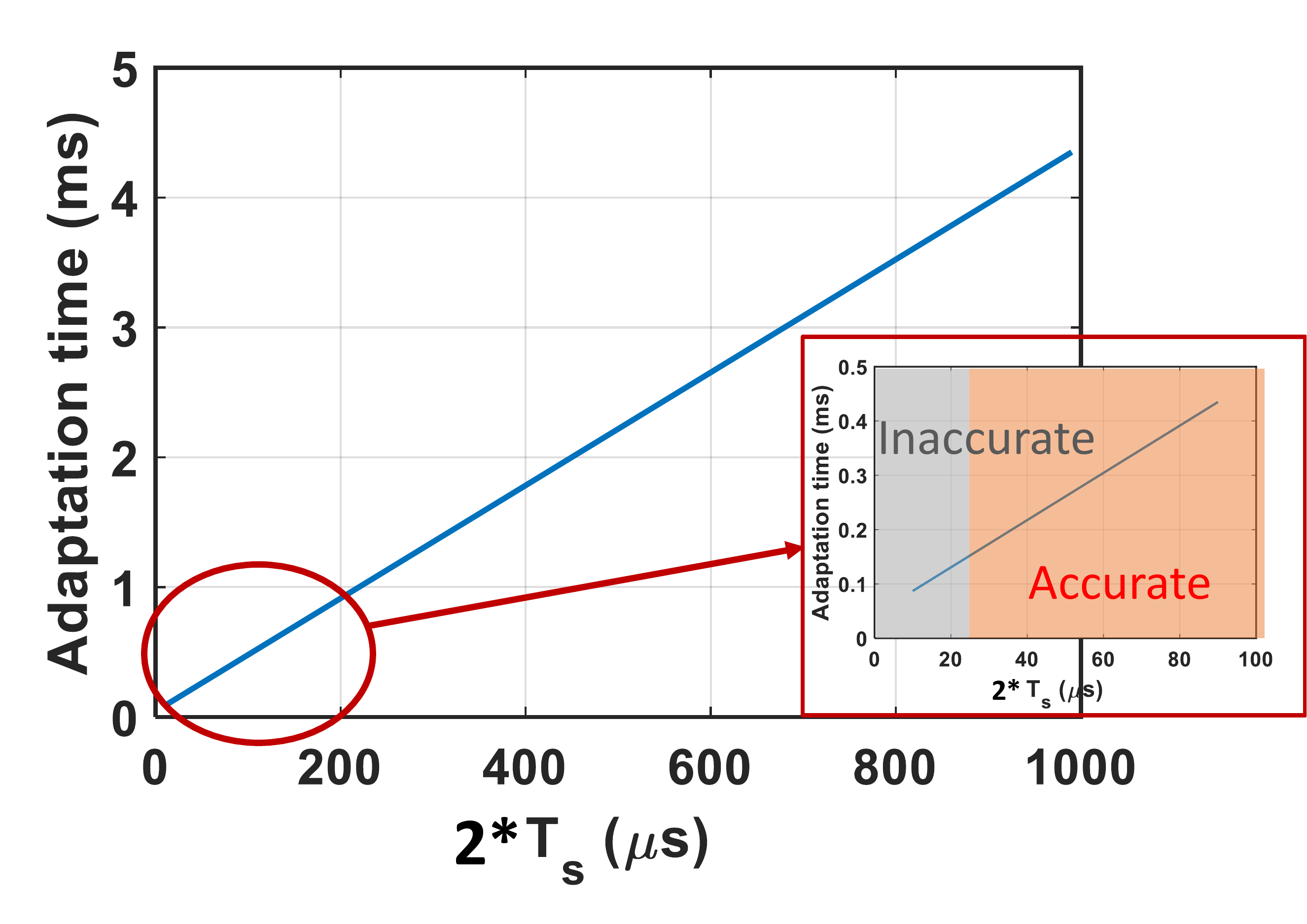}
\caption{Adaptation time vs. ADC sampling time (T\textsubscript{s}) with K\textsubscript{V\textsubscript{G}}K\textsubscript{G}=0.6 and assuming T\textsubscript{V\textsubscript{G}}$\approx$ 10$\mu$s.}
% \vspace{-1.5em}
\label{fig:controladaptationtime}
\end{figure}

Some known timing characteristics are: \begin{itemize}
    \item T\textsubscript{V\textsubscript{G}} $\approx$4 - 10 $\mu$s per step depending on the step size
    \item T\textsubscript{process} is negligible in the ranges of <$\mu$s
    \item T\textsubscript{s}$\approx$2 $\mu$s - 1 ms or F\textsubscript{s}$\approx$1 - 500 kHz
    \item T\textsubscript{Load}$\approx$ T\textsubscript{V\textsubscript{G}}$\approx$4 - 10 $\mu$s or 
    F\textsubscript{Load}$\approx$ F\textsubscript{V\textsubscript{G}}$\approx$100 - 250 kHz.
\end{itemize}  
Since highest F\textsubscript{s} is 500 kHz, F\textsubscript{s} << F\textsubscript{Load}, a linear increase of adaptation time vs. ADC sampling time is observed in Fig. \ref{fig:controladaptationtime}, assuming T\textsubscript{V\textsubscript{G}}$\approx$ 10$\mu$s and a negligible processing time with K\textsubscript{V\textsubscript{G}}K\textsubscript{G}=0.6. Note that in the figure, a sufficient margin higher than T\textsubscript{V\textsubscript{G}} ( >25us of total ADC measurement time for 2 samples) is needed for an accurate reading. The simulations in Fig. \ref{fig:controladaptationtime} allow for a more accurate estimation of the minimum adaptation time for the control loop. The adaptation time is strongly dependent on the ADC sampling rate, so if ADC sampling rate can be increased, the adaptation time can be reduced. Otherwise, the minimum adaption time is about 150 $\mu$s with the current sampling rate with the number of steps to settling being approximately four.

\section{Measurement Results}
\label{sec:results}

% \begin{figure}
% \centering
% \includegraphics[width=85mm]{figures/3gfffbtrans.pdf}
% \caption{CHANGE!!!!!! 3 GHz Data}
% \vspace{-1.5em}
% \label{fig:3gfffbtrans}
% \end{figure}

% \begin{figure}
% \centering
% \includegraphics[width=85mm]{figures/2.5GHzfbonlytrans.pdf}
% \caption{CHANGE!!!!!! 3 GHz Data}
% \vspace{-1.5em}
% \label{fig:2.5gtrans}
% \end{figure}

\subsection{Results from Feedback Only Control}

{% <-- We enclose the table in a group so that any redefinitions
%% are automatically undone at the end of the group.
%
\setlength{\tabcolsep}{1mm}%
\renewcommand{\arraystretch}{1.2}% for the vertical padding of table cells
\newcommand{\CPcolumnonewidth}{15mm}%
\newcommand{\CPcolumntwowidth}{32mm}%
\newcommand{\CPcolumnthreewidth}{35mm}%
\newcommand{\CPcolumnfourwidth}{2mm}%
\begin{table}[h]
\caption{Design Comparison}
\small% IMS: need this to get the 9pt text size in table cells % TODO is correct ?
\centering
\begin{tabular}{|l|l|l|}\hline
%\multirow{2}{6mm}{\parbox{8mm}{{\bfseries Font Size}}} & \multicolumn{3}{c|}{\raisebox{-0.25mm}{\bfseries Appearance (in Times New Roman or Times)}}\\ \cline{2-4}
\raisebox{-0.5mm}{\bfseries Control} & \raisebox{-0.5mm}{\bfseries Interference On Off} & \raisebox{-0.5mm}{\bfseries Interference Level Change}\\ \hline
\parbox[t]{\CPcolumnonewidth}{\strut Incremental\\ Adaptation \strut}& \parbox[t]{\CPcolumntwowidth}{\strut -Slowest adaptation time ($\sim$600 $\mu$s) \\-wider frequency range adaptation \strut}& \parbox[t]{\CPcolumnthreewidth}{\strut -skips steps during interference level change \\-settles to different V\textsubscript{G} \strut} \\ \hline
\parbox[t]{\CPcolumnonewidth}{\strut Look-Up \\Table\strut}& \parbox[t]{\CPcolumntwowidth}{\strut -Fastest adaptation time ($\sim$180 $\mu$s) \\-all steps are visible during interference level change \\-narrower frequency  range adaptation \\-requires more memory space to implement full look-up table \strut}& \parbox[t]{\CPcolumnthreewidth}{\strut -all steps are visible during interference level change \\-settles to the same V\textsubscript{G} most of the time \strut} \\ \hline
\parbox[t]{\CPcolumnonewidth}{\strut One-Shot \\+ \\Incremental \\Adaptation\strut}& \parbox[t]{\CPcolumntwowidth}{\strut -Faster adaptation time ($\sim$450 $\mu$s) \\-wider frequency range adaptation \\-frequency specific one-shot or underestimation during one-shot\strut}& \parbox[t]{\CPcolumnthreewidth}{\strut -skips steps during interference level change \\-frequency specific one-shot or underestimation during one-shot\strut} \\ \hline
%12 & \parbox[t]{\CPcolumntwowidth}{{author name},\\author affiliation,\\email address\strut} & \\ \hline
%18 & title & \\ \hline
\end{tabular}
\vspace{0 em}
\label{tab:controlcomp}
\end{table}
}% end of group enclosing the table

\begin{figure}
\centering
\includegraphics[width=85mm]{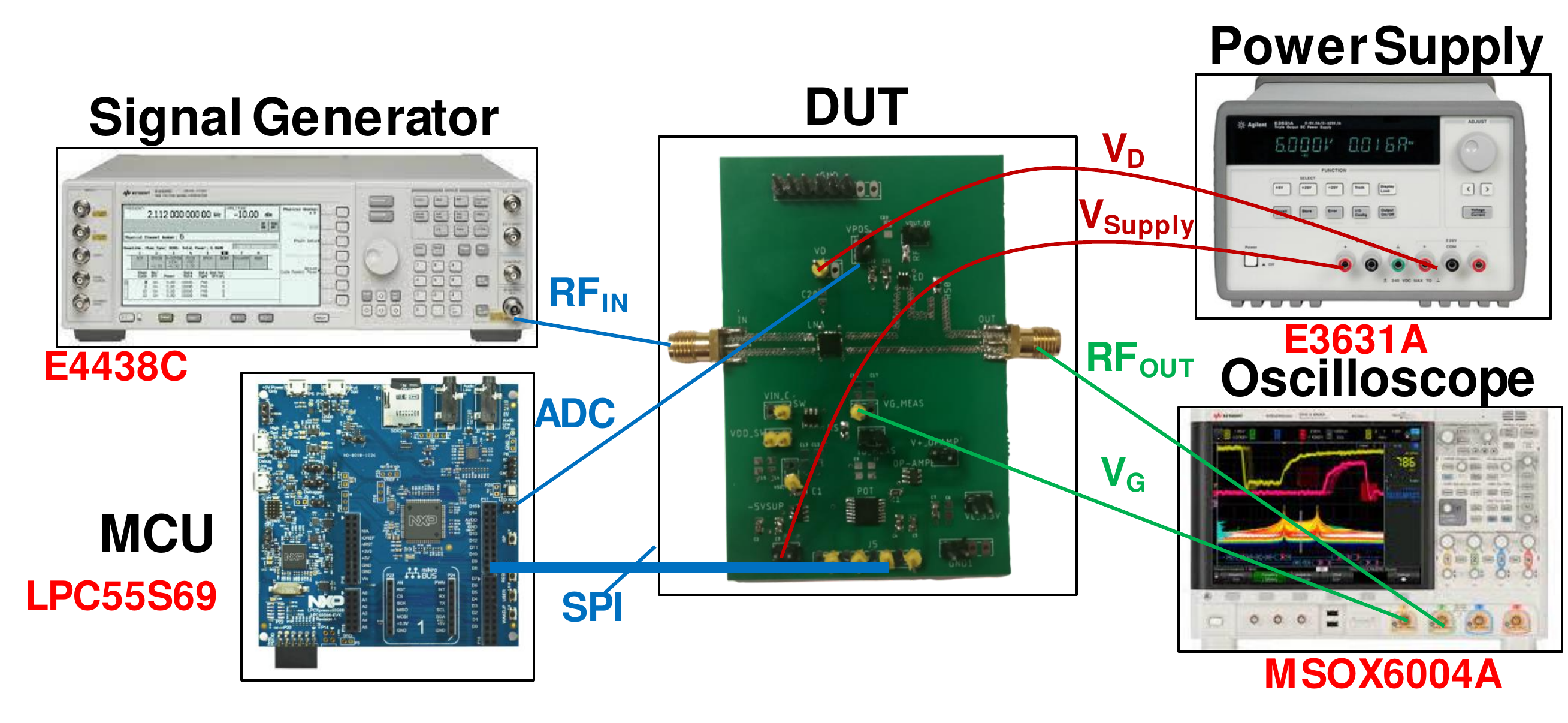}
\caption{Measurement setup.}
% \vspace{-1.5em}
\label{fig:measurementsetup}
\end{figure}

\begin{figure*}
\centering
\includegraphics[width=180mm]{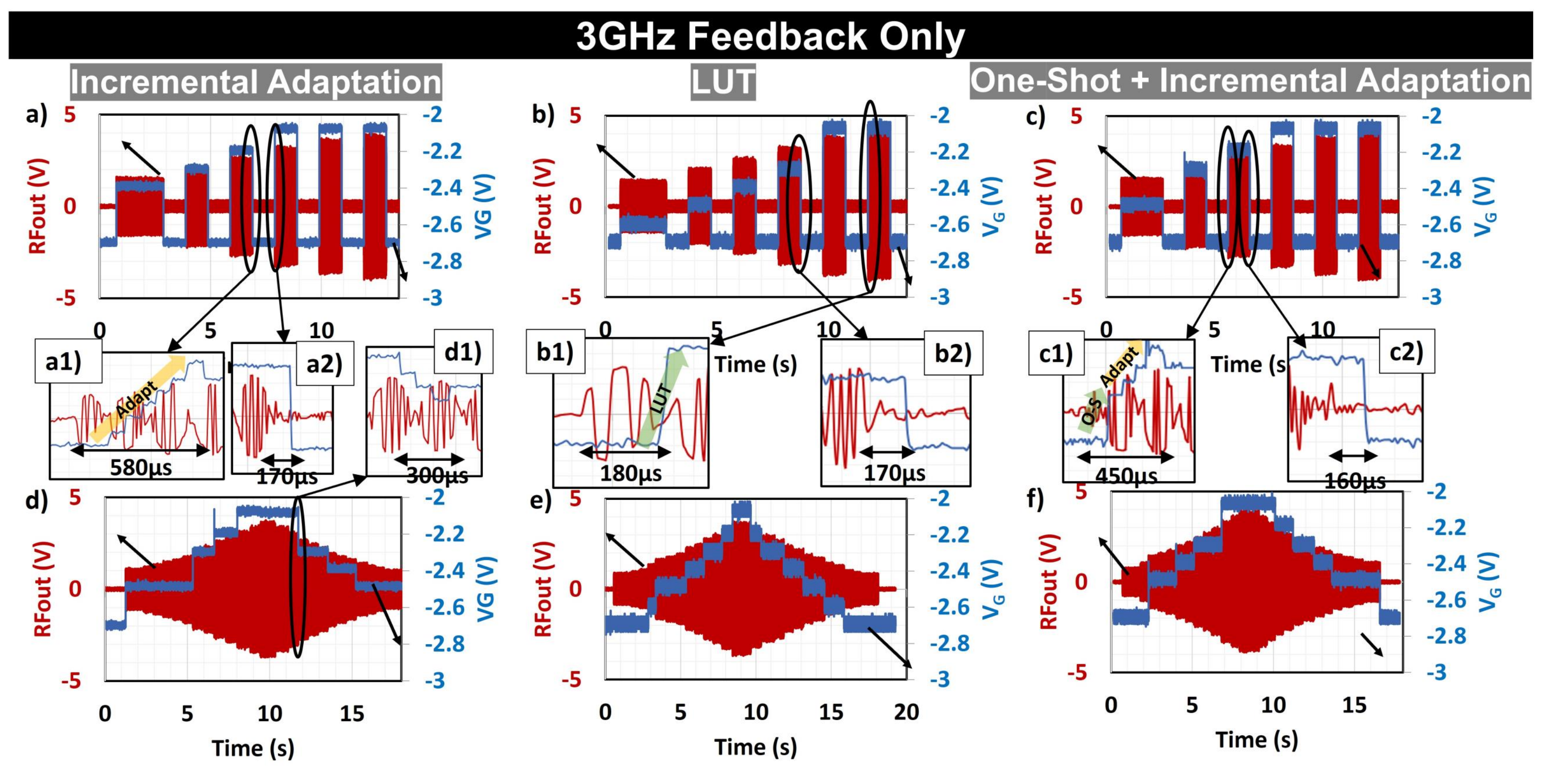}
\caption{3 GHz transient data with V\textsubscript{D} = 10 V. Transient data of RF output and V\textsubscript{G} for input interference sequence of -9.5, -22.5, -6.5, -22.5, -4.5, -22.5, -2.5, -22.5, -1, -22.5, 0, -22.5 dBm with control a) incremental adaptation only, b) LUT, c) one-shot + incremental adaptation. a1)-a2), b1)-b2) and c1)-c2) tuning transient for the appearance of interference. 3 GHz transient data with input interference from -12.5 dBm to -0.5 dBm and back down to -12.5 dBm with control d) incremental adaptation only, e) LUT, f) one-shot + incremental adaptation. d1) Tuning characteristics for decreasing interference for incremental adaptation control. }
\vspace{-1.5em}
\label{fig:3gfbonlytrans}
\end{figure*}

Feedback-only control implements a temporal control where the tuning uses two samples of the output signal levels at different times. The results are summarized in Table \ref{tab:controlcomp}.

\begin{figure}
\centering
\includegraphics[width=75mm]{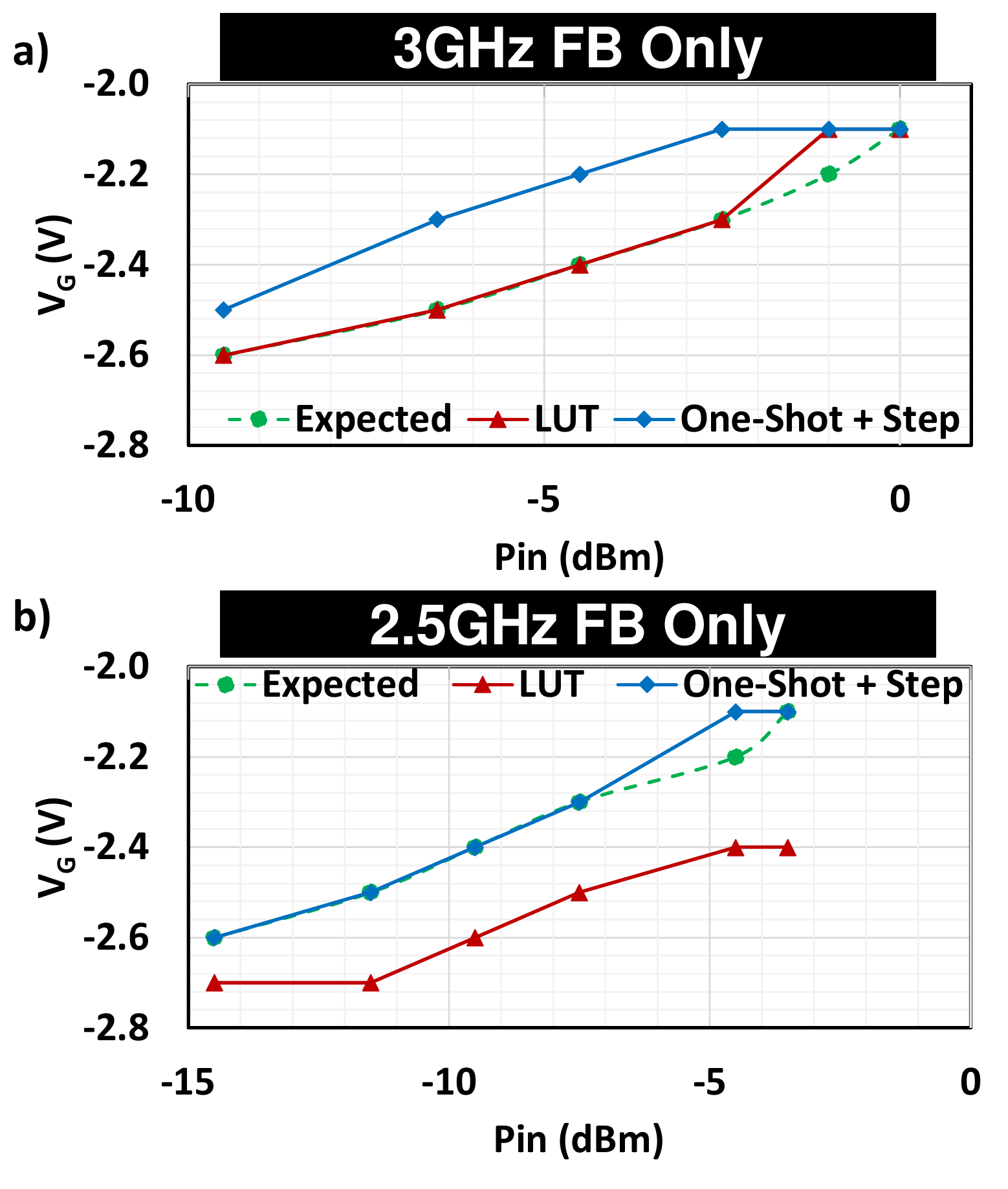}
\caption{a) Settled V\textsubscript{G} value for different adaptation methods at 3GHz on feedback-only board; b) settled V\textsubscript{G} value for different adaptation methods at 2.5 GHz on feedback-only board.}
\vspace{-1.5em}
\label{fig:2.5g3gcomp}
\end{figure}

Fig. \ref{fig:3gfbonlytrans} presents the adaptation and timing characteristics for different control methods with the measurement setup in Fig. \ref{fig:measurementsetup}. Note that after zooming into the transient, the RF output measured using an oscilloscope is not a perfect sine wave due to the undersampling of the oscilloscope trying to capture the signal with a larger time scale. Fig. \ref{fig:3gfbonlytrans}a) - c) show the adaptation of the LNA by varying different interference levels on and off. Different high interference levels at 3 GHz are forced at the input with the expectation of V\textsubscript{G} to increase from -2.6 V to -2.1 V in the increment of 0.1 V; at low interference levels, V\textsubscript{G} is expected to drop down to -2.7 V. A summary of different settled V\textsubscript{G} with respect to different input interference levels is presented in Fig. \ref{fig:2.5g3gcomp}a). As expected, the LUT(Fig. \ref{fig:3gfbonlytrans}b) is more aligned with the expected V\textsubscript{G} as the LUT values are specifically for 3 GHz while both the incremental adaptation only(Fig. \ref{fig:3gfbonlytrans}a) and the one-shot + incremental adaptation(Fig. \ref{fig:3gfbonlytrans}c) overestimates V\textsubscript{G}. Fig. \ref{fig:3gfbonlytrans}a1), b1) and c1) show the tuning times to adapt to an interference level for the incremental adaptation, LUT and one-shot + incremental adaptation are 580, 180 and 450 $\mu$s respectively. Fig. \ref{fig:3gfbonlytrans}a2), b2) and c2) show the tuning times to adapt to a disappearing interference for the three control methods are very similar, around 170 $\mu$s. Fig. Fig. \ref{fig:3gfbonlytrans}d) - f) show the adaptation of the LNA when the interference level increases from -12.5 dBm to -0.5 dBm and backs down to -12.5 dBm in increments of 1 dBm. With the LUT method in Fig. \ref{fig:3gfbonlytrans}e), every step of V\textsubscript{G} is shown and roughly the same settling V\textsubscript{G} for the same interference level, whereas for the incremental adaptation (Fig. \ref{fig:3gfbonlytrans}d) and one-shot + incremental adaptation method(Fig. \ref{fig:3gfbonlytrans}f), some V\textsubscript{G} steps are skipped and sometimes different V\textsubscript{G} values for the same interference level. Fig. \ref{fig:3gfbonlytrans}d1) shows that when the interference level decreases, the control loop is able to step down and adapt.

\begin{figure}
\centering
\includegraphics[width=75mm]{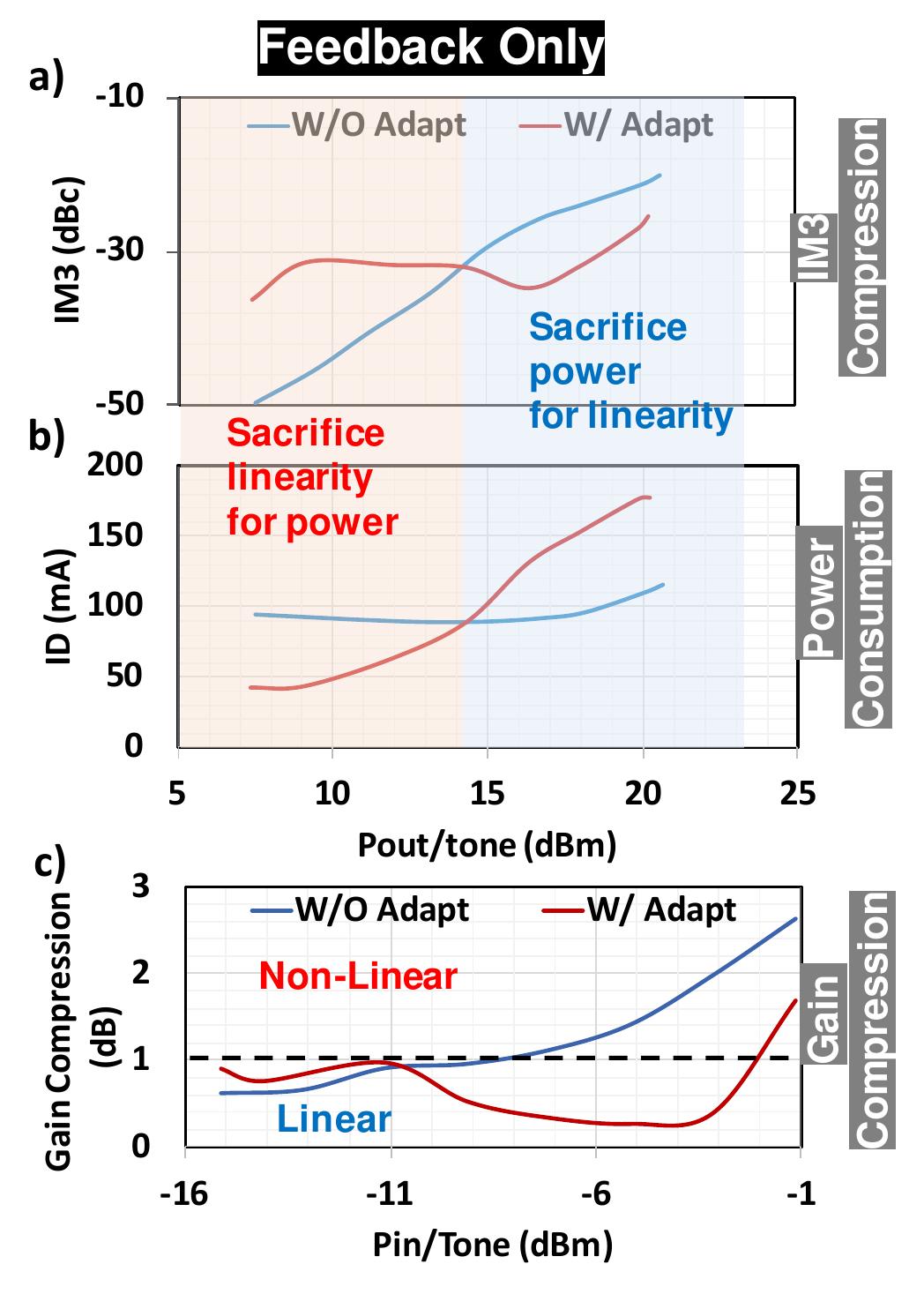}
\caption{3 GHz data for feedback board comparison with and without adaption for two-tone measurements for a) IM3 compression, b) drain current of the LNA and c) gain compression.}
\vspace{-1.5em}
\label{fig:linearityvspower}
\end{figure}

\begin{figure}
\centering
\includegraphics[width=75mm]{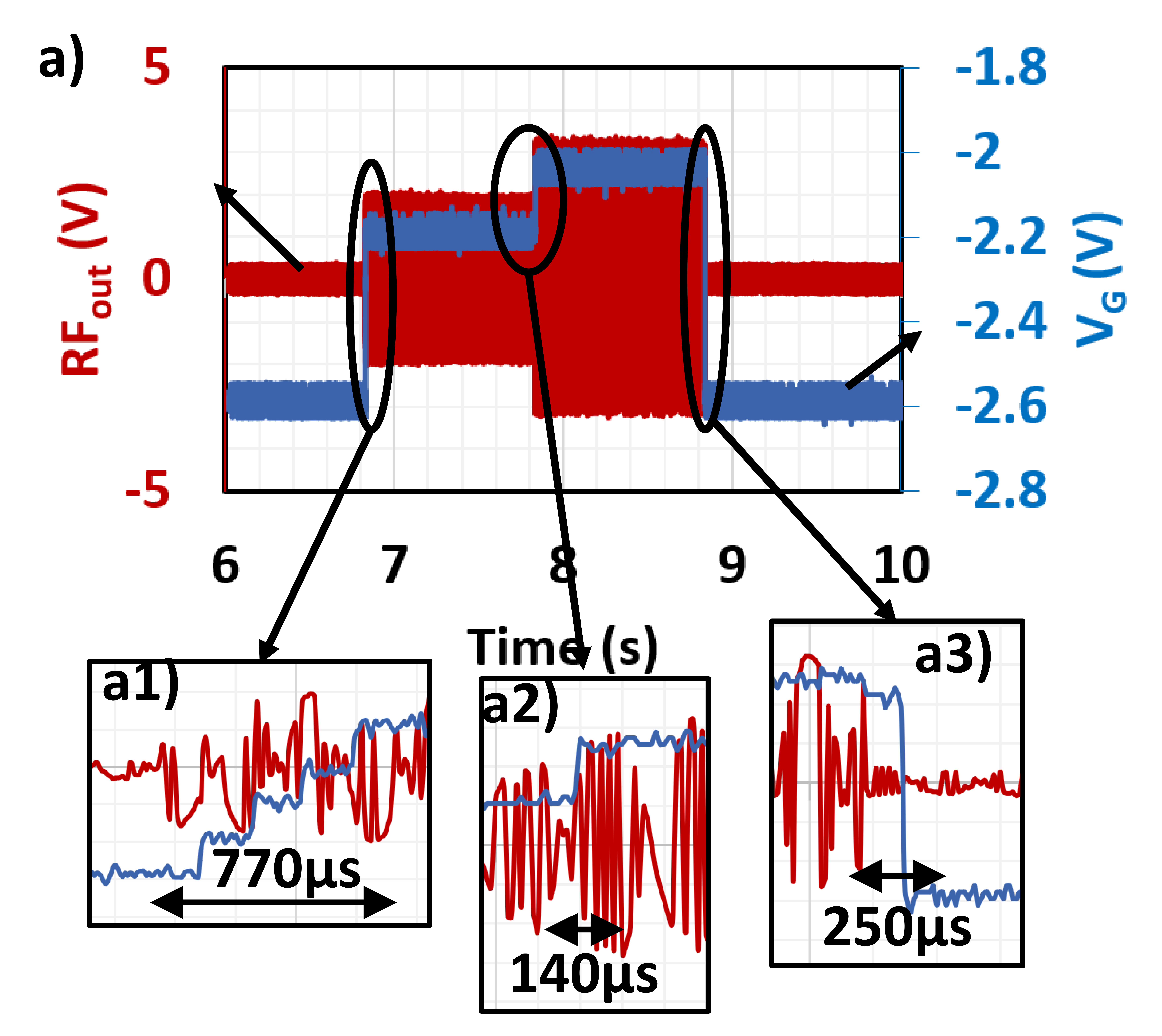}
\caption{3 GHz adaptation data for feedforward + feedback board when input interference level changes from -5.5 dBm to -2.5 dBm. a) shows the overall tuning of V\textsubscript{G} and the effect on the output. a1)-a3) show timing characteristics for different tuning steps.}
\vspace{-1.0em}
\label{fig:fffbcontrol}
\end{figure}

Fig. \ref{fig:2.5g3gcomp} shows the settling V\textsubscript{G} values for different methods of control vs. input power with frequencies of 3 GHz and 2.5 GHz. In the 3 GHz plot in Fig. \ref{fig:2.5g3gcomp}a), LUT method almost matches up with all the expected V\textsubscript{G} to bring the LNA back to linearity whereas the incremental adaptation only and  one-shot + incremental adaptation overestimates for many P\textsubscript{IN} values. However, in the 2.5 GHz plot in Fig. \ref{fig:2.5g3gcomp}b), LUT significantly underestimates the necessary V\textsubscript{G} for linearity due to the LUT is only captured at 3 GHz and the algorithm tries to match the 2.5 GHz ADC values with the 3 GHz values. On the contrary, the one-shot + incremental adaptation method matches up with the expected better at 2.5 GHz than at 3 GHz due to the set threshold value being better suited at 2.5 GHz since the threshold value is chosen to adapt to a wider range of frequencies.

Fig. \ref{fig:linearityvspower} shows the trade-offs between an adaptive system vs. a non-adaptive system operating in nominal conditions with a two-tone measurement. Fig. \ref{fig:linearityvspower}a) and b) show that when the interference is low, the adaptive system consumes less power while still maintaining linearity for the LNA in comparison to the non-adaptive system having a higher IM3 compression and consumes more power; when the interference is high, the adaptive system consumes more power to bring the LNA back to linearity with a higher IM3 compression compared to the non-adaptive system with lower IM3 compression (LNA is non-linear) and lower power consumption. Fig. \ref{fig:linearityvspower}c) shows that the adaptive system is able to keep the LNA in the linearity range over a wider input range than the non-adaptive system in the nominal condition.

\subsection{Results from Feedforward + Feedback Control}
Feedforward + feedback design allows a spacial control of the system where both the input and output signal levels can be sampled, and the linearity is determined by comparing the measured gain with the expected gain to ensure no gain compression. Fig. \ref{fig:fffbcontrol} shows that the tuning circuit is able to use spatial control to tune the V\textsubscript{G} of the LNA as the interferences appear, increase and disappear; however, this method also suffers from an overestimation of the V\textsubscript{G} value that a higher V\textsubscript{G} value is determined. Note from Fig. \ref{fig:fffbcontrol}a1), unlike the overshoot in the feedback only incremental adaptation control, since this board has the extra degree of observability and that the measured gain is directly compared with the expected gain, tuning does not give overshoot in V\textsubscript{G}. From Fig. \ref{fig:fffbcontrol}a1-3), the tuning times for the interference appearance, increasing of interference, and disappearance of interference are 770 $\mu$s, 140 $\mu$s and 250 $\mu$s respectively. The tuning times are below 1 ms; however, as shown in Fig. \ref{fig:fffbcontrol}a1), a longer tuning time is needed than the feedback only incremental adaptation control due to ADC sampling time is lengthened to accurately measure the input power being close to the sensitivity level.       

\subsection{System Comparison}
\label{sec:syscomparison}
\begin{figure*}
\centering
\includegraphics[width=180mm]{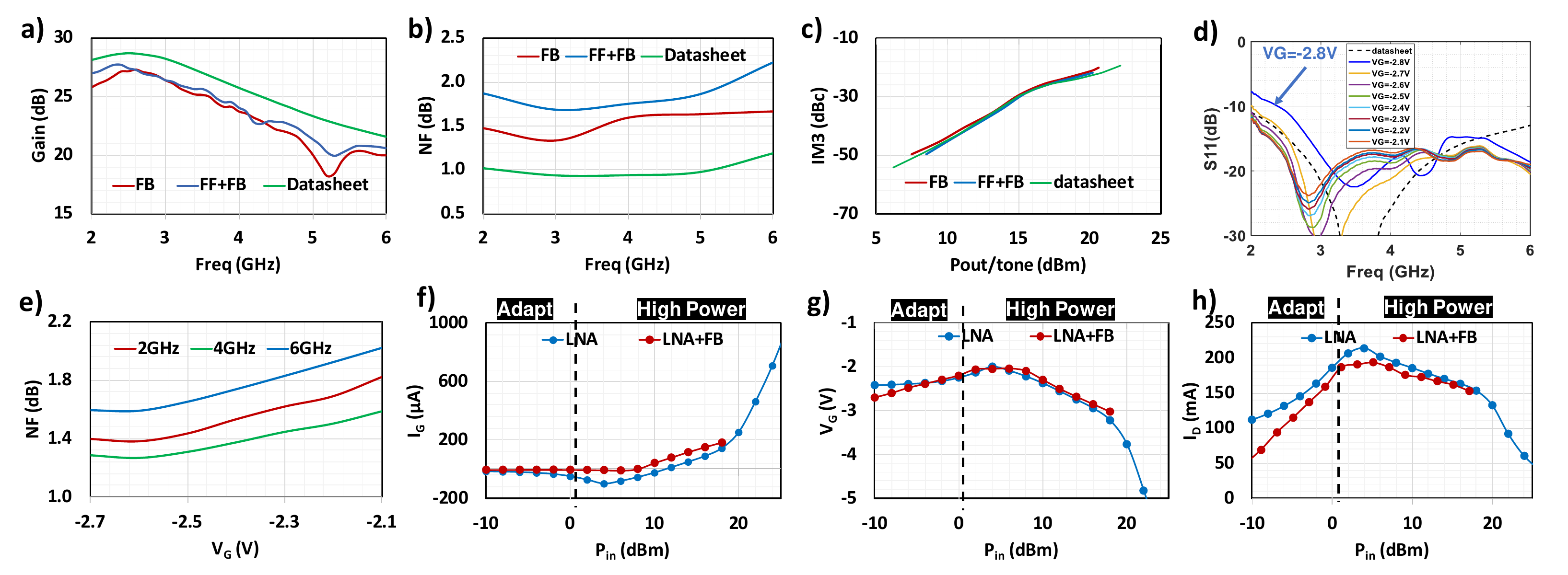}
% \caption{a)-c) shows LNA characteristic comparison with feedback only (FB) board, feedforward and feedback (FF+FB) board and datasheet values at V\textsubscript{D} = 10 V, I\textsubscript{D} $\approx$ 100 mA a) gain vs. frequency, b) noise figure (NF) vs. frequency, c) IM3 compression vs. P\textsubscript{OUT} for each tone with 10 MHz spacing at 4 GHz. d) Return loss (S11) for the FB board with different V\textsubscript{G} to the LNA.  }
\caption{Different LNA characteristics such as gain, NF, and IM3 compression are compared between the feedback system, feedback + feedforward system, and the datasheet in a)-c) respectively (V\textsubscript{D} = 10 V, I\textsubscript{D} $\approx$ 100 mA).  d) shows return loss (S11) for the feedback system. e) shows NF vs. V\textsubscript{G} for different frequencies. f)-h) shows the high input power comparison between LNA with a series resistance of 5.03 k$\Omega$ without adaptation (Fig. \ref{fig:VGHighPower}) and LNA in the feedback system at 3 GHz.}
\vspace{-1.5em}
\label{fig:meascomp} 
\end{figure*}

Fig. \ref{fig:meascomp} presents different LNA characteristics after implementation on the feedback-only board and the combined feedforward and feedback board with the datasheet values in similar bias conditions of V\textsubscript{D} = 10 V and I\textsubscript{D} $\approx$ 100 mA. Note that the datasheet values are measured directly on the die, by implementing on a PCB, some degree of degradation in the gain and NF are expected. Fig. \ref{fig:meascomp}a) shows the gain comparison over the frequency range of 2-6 GHz. The gain of the boards shows a few dB of degradation in comparison to the datasheet while little difference between the different boards. Fig. \ref{fig:meascomp}b) shows the noise figure (NF) vs. frequency for different boards. The feedback-only board has an NF about 0.5 dB higher than the datasheet, and as expected with the feedforward and feedback combined board, the NF is again about 0.3 dB higher than the feedback-only layout due to the extra directional coupler before the LNA. Fig. \ref{fig:meascomp}c) shows the 3rd order intermodular product (IM3) compression vs. desired output power for each tone. The IM3 compression are relatively close across different boards and the datasheet.

Fig. \ref{fig:meascomp}d) shows different return loss (S11) with respect to different V\textsubscript{G} values. As shown in the plot, when V\textsubscript{G}=-2.8 V, the return loss is worse than all of the other voltages, so when implementing the control loop, only V\textsubscript{G} greater than -2.8 V is considered. For V\textsubscript{G} between -2.7 V to -2.1 V, the majority of the responses have an S11 lower than -15 dBm, other parts have an S11 lower than -10 dBm. Fig. \ref{fig:meascomp}e) shows NF across V\textsubscript{G} tuning range for 2, 4 and 6 GHz. Across V\textsubscript{G} the NF can increase about 0.4 dB. When interference is low, instead of operating at nominal V\textsubscript{G} of -2.4 V with a higher NF, lower NF can be achieved in the adaptive system with V\textsubscript{G} of -2.7 V. When interference is high, higher linearity is achieved with higher NF and power consumption.   

Fig. \ref{fig:meascomp}f)-h) shows the comparison of the high power data for the LNA in Fig. \ref{fig:VGHighPower} and LNA in a feedback system. In the adaptation range, the V\textsubscript{G} in the feedback system is continuously increasing to maintain linearity by adapting to the current input while the V\textsubscript{G} increase in the LNA only system is caused by the non-linearity of the LNA. The difference in linearity in the two systems can also be observed in the I\textsubscript{G}, where the linear system has a consistent I\textsubscript{G} in the ranges of $\mu$A whereas the non-linear system sees more of a revered biased current. I\textsubscript{D} for both systems show an increase because of the increase in V\textsubscript{G}. In the high power range, the currents and the voltages are starting to overlap showing similar high power effects as described in Sec. \ref{sec:Hpower}.

Table \ref{tab:designLNAcomp} shows comparisons between the datasheet LNA in nominal condition, LNA with feedback control and LNA with feedback + feedforward control. According to the datasheet \cite{LNA}, the GaN LNA's nominal condition is V\textsubscript{D}=10 V and I\textsubscript{D}=100 mA, which consumes about 1 W of power. The controls consume $\leq$10$\%$ of total LNA nominal power for a better linearity and LNA power consumption trade-off. Most of the control power derives from the microcontroller consuming 80 mW of power, which is about 80-90$\%$ of total control power. LNA with control gives a wide tuning range in power and linearity. When the system is operating in low-power mode in a large EW receiver array, the power saving can be significant. Extra system power is consumed when higher linearity is required to decode the signal.

{% <-- We enclose the table in a group so that any redefinitions
%% are automatically undone at the end of the group.
%
\setlength{\tabcolsep}{1mm}%
\renewcommand{\arraystretch}{1.2}% for the vertical padding of table cells
\newcommand{\CPcolumnonewidth}{25mm}%
\newcommand{\CPcolumntwowidth}{15mm}%
\newcommand{\CPcolumnthreewidth}{20mm}%
\newcommand{\CPcolumnfourwidth}{20mm}%
\begin{table}[h]
\caption{Performance Comparison}
\small% IMS: need this to get the 9pt text size in table cells % TODO is correct ?
\centering
\begin{tabular}{|l|l|l|l|}\hline
%\multirow{2}{6mm}{\parbox{8mm}{{\bfseries Font Size}}} & \multicolumn{3}{c|}{\raisebox{-0.25mm}{\bfseries Appearance (in Times New Roman or Times)}}\\ \cline{2-4}
\raisebox{-0.25mm}{\bfseries Specs } &
\raisebox{-0.25mm}{\bfseries LNA} & \raisebox{-0.25mm}{\bfseries LNA+FB} & \raisebox{-0.25mm}{\bfseries LNA+FB+FF}\\ \hline
\parbox[t]{\CPcolumnonewidth}{\strut \textbf{Gain (dB)$^*$}\strut} &
\parbox[t]{\CPcolumntwowidth}{\strut 28 \strut} & \parbox[t]{\CPcolumnthreewidth}{\strut 26.5\strut} & \parbox[t]{\CPcolumnfourwidth}{\strut 26.5\strut}\\ \hline
\parbox[t]{\CPcolumnonewidth}{\strut \textbf{NF (dB)$^*$}\strut} &
\parbox[t]{\CPcolumntwowidth}{\strut 1 \strut} & \parbox[t]{\CPcolumnthreewidth}{\strut 1.4\strut} & \parbox[t]{\CPcolumnfourwidth}{\strut 1.7\strut}\\ \hline
\parbox[t]{\CPcolumnonewidth}{\strut \textbf{P\textsubscript{1dB,IN} (dBm)}\strut} &
\parbox[t]{\CPcolumntwowidth}{\strut -7 \strut} & \parbox[t]{\CPcolumnthreewidth}{\strut -10.5 $\sim$ 0.5 \strut} & \parbox[t]{\CPcolumnfourwidth}{\strut -14 $\sim$ 0.5\strut}\\ \hline
\parbox[t]{\CPcolumnonewidth}{\strut \textbf{LNA Power (W)}\strut} &
\parbox[t]{\CPcolumntwowidth}{\strut 1 \strut} & \parbox[t]{\CPcolumnthreewidth}{\strut 0.5 $\sim$ 2 \strut} & \parbox[t]{\CPcolumnfourwidth}{\strut 0.3 $\sim$ 1.8 \strut}\\ \hline
\parbox[t]{\CPcolumnonewidth}{\strut \textbf{Control Power (W)}\strut} &
\parbox[t]{\CPcolumntwowidth}{\strut 0 \strut} & \parbox[t]{\CPcolumnthreewidth}{\strut 0.09\strut} & \parbox[t]{\CPcolumnfourwidth}{\strut 0.1\strut}\\ \hline
\parbox[t]{\CPcolumnonewidth}{\strut \textbf{Observation Points}\strut} &
\parbox[t]{\CPcolumntwowidth}{\strut 0 \strut} & \parbox[t]{\CPcolumnthreewidth}{\strut 1\strut} & \parbox[t]{\CPcolumnfourwidth}{\strut 2\strut}\\ \hline
\parbox[t]{\CPcolumnonewidth}{\strut \textbf{Control Method}\strut} &
\parbox[t]{\CPcolumntwowidth}{\strut none \strut} & \parbox[t]{\CPcolumnthreewidth}{\strut Temporal\strut} & \parbox[t]{\CPcolumnfourwidth}{\strut Spacial\strut}\\ \hline
\parbox[t]{\CPcolumnonewidth}{\strut \textbf{Tuning Step (us)}\strut} &
\parbox[t]{\CPcolumntwowidth}{\strut none \strut} & \parbox[t]{\CPcolumnthreewidth}{\strut 85\strut} & \parbox[t]{\CPcolumnfourwidth}{\strut 150\strut}\\ \hline

%12 & \parbox[t]{\CPcolumntwowidth}{{author name},\\author affiliation,\\email address\strut} & \\ \hline
%18 & title & \\ \hline
\end{tabular}
% \vspace{+0.5 em}
\footnotesize{$^*$ 3 GHz Data}
\label{tab:designLNAcomp}

\end{table}

% \vspace{-1 em}

}% end of group enclosing the table

Feedforward + feedback design includes two pairs of directional coupler and ED. With the extra directional coupler and ED1 before the LNA, NF increases by $\approx$ 0.3 dB, and overall control power consumption increases by another 10 mW than feedback only design; however, the design gives more information on the input signal that will be more relevant in future works of an adaptive receiver. For example, if a filter was placed to remove the interference signal, the extra information on the input signal would allow us to determine if the interference is still present or has already been removed by the filter. The feedback-only design provides a simpler solution for the current application to detect the presence and the level of interference signal at the output of the LNA. Contrary to the spacial control in the feedforward + feedback design, feedback-only design implements a temporal control. At the current stage, the extra degree of observability at the input of the LNA is not needed as no filter is present.

\begin{table*}
\begin{center}
  \caption{Comparison Table}
  \label{tbl:compare}
  \includegraphics[width=180mm]{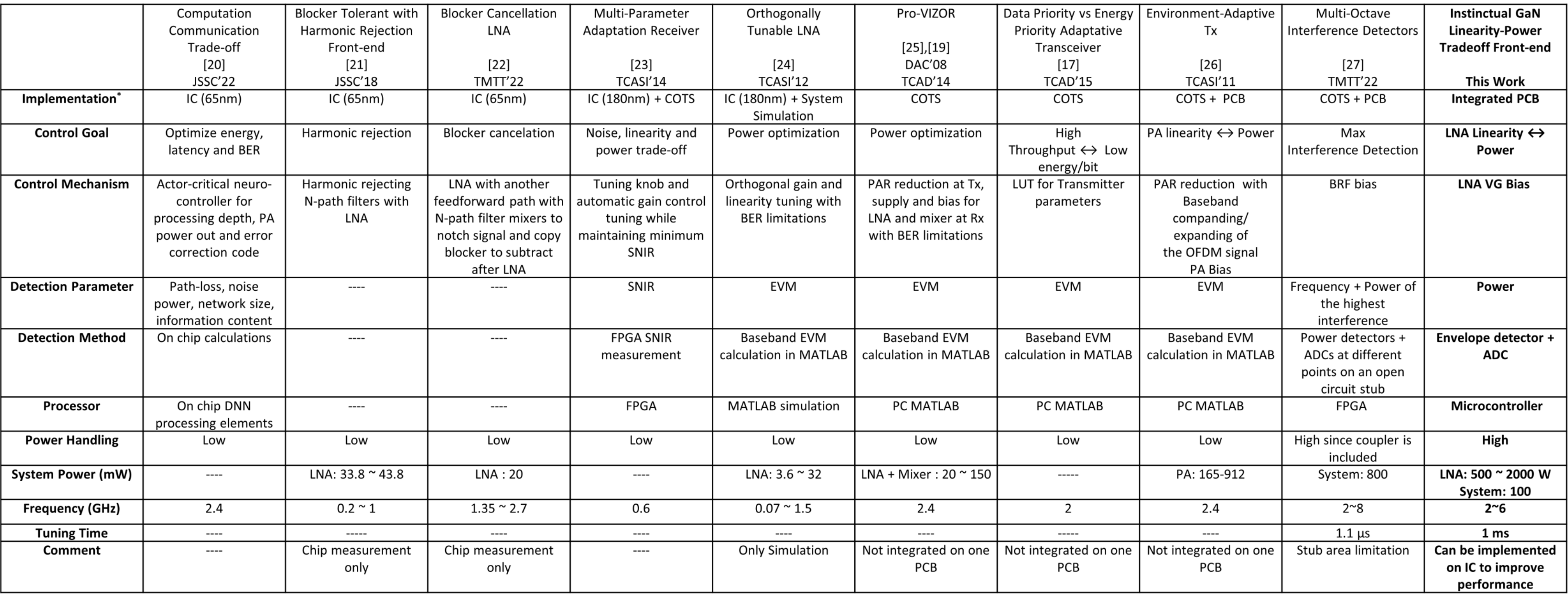}
\end{center}
% \vspace{-1.5em}
\footnotesize{   $^*   $  IC+COTS: Custom IC with COTS connected through SMA connector. COTS: COTS connected through SMA connector. COTS+PCB: some custom microwave components connected with some components integrated into the PCB. Integrated PCB: components are integrated into one PCB without any SMA connections to other components.}
\end{table*}

\subsection{Comparison Table}

Table \ref{tbl:compare} shows the different works that contribute to adaptive transceivers over different implementation methods for higher resilience towards undesired interference.

\section{Future Work}
\label{sec:future}

The current work can be further extended by implementing a complete RF receiver front end, adding interference frequency detection and/or interference filtering circuitry for interference compression, and better use of some frequency-dependent control methods. The interference frequency detection circuits can assist with deciding other signals for interference cases such as strong signal and strong interference. The adaptation time as well as the power consumption of the controlling circuit can be further minimized. If Apollo4 microcontroller is utilized, the microcontroller would only consume mW's of power and reduced the total control power to <3$\%$ of nominal LNA power consumption \cite{ap4}. A different LNA with orthogonal tunability can be explored so that with a blocker, the gain can also be tuned to achieve large signal linearity. Noise figure of the front end can also be further improved.

The concept of the work can be expanded into other technologies by having a strong correlation between the bias voltages of the LNA and linearity and the accessibility of the bias voltages. The work can also be expanded into other frequencies with correct characterizations of different components. The LNA and the onboard directional couplers are specifically for 2-6 GHz, and the envelope detector 2 can function up to 43.5 GHz.  

If similar systems were to be built in low-power receivers, a custom ASIC design can bring down the control loop power by orders of magnitude. Such systems are part of future work and will expand the applicability of instinctual receivers to much wider power categories of RF systems.

\section{Conclusion}

This paper presents the first instinctual GaN LNA system demonstration with intelligent localized sensing, processing, and feedback controls to achieve sub-1ms adaptation to a variety of interference scenarios. GaN LNA is utilized for the high power handling capabilities in radar and EW applications. The system consumes $\leq$10$\%$ of nominal LNA power to provide a wide range of tuning. The linearity tuning range is about 11 dB; LNA power consumption tuning range is about 0.5-2 W; NF changes about 0.4 dB across the tuning range. When the frequency is known using LUT, the system adapts to an interference very accurately to bring the LNA back to linearity. When the frequency is unknown, the system is still able to adapt to interference with extra power consumption using either incremental adaptation only or one-shot + incremental adaptation. The feedback-only board provides simplicity of the design for this application. The adaptation time for the system is <1 ms which closely matches the theoretical simulations.

% if have a single appendix:
%\appendix[Proof of the Zonklar Equations]
% or
%\appendix  % for no appendix heading
% do not use \section anymore after \appendix, only \section*
% is possibly needed

% use appendices with more than one appendix
% then use \section to start each appendix
% you must declare a \section before using any
% \subsection or using \label (\appendices by itself
% starts a section numbered zero.)
%

% \appendices
% \section{Proof of the First Zonklar Equation}
% Appendix one text goes here.

% % you can choose not to have a title for an appendix
% % if you want by leaving the argument blank
% \section{}
% Appendix two text goes here.

% use section* for acknowledgment
% \section*{Acknowledgment}

% The authors would like to thank...

% Can use something like this to put references on a page
% by themselves when using endfloat and the captionsoff option.
\ifCLASSOPTIONcaptionsoff
  \newpage
\fi

% trigger a \newpage just before the given reference
% number - used to balance the columns on the last page
% adjust value as needed - may need to be readjusted if
% the document is modified later
%\IEEEtriggeratref{8}
% The "triggered" command can be changed if desired:
%\IEEEtriggercmd{\enlargethispage{-5in}}

% references section

% can use a bibliography generated by BibTeX as a .bbl file
% BibTeX documentation can be easily obtained at:
% http://mirror.ctan.org/biblio/bibtex/contrib/doc/
% The IEEEtran BibTeX style support page is at:
% http://www.michaelshell.org/tex/ieeetran/bibtex/
%\bibliographystyle{IEEEtran}
% argument is your BibTeX string definitions and bibliography database(s)
%\bibliography{IEEEabrv,../bib/paper}
%
% <OR> manually copy in the resultant .bbl file
% set second argument of \begin to the number of references
% (used to reserve space for the reference number labels box)
% \begin{thebibliography}{1}

% \bibitem{IEEEhowto:kopka}
% H.~Kopka and P.~W. Daly, \emph{A Guide to \LaTeX}, 3rd~ed.\hskip 1em plus
%   0.5em minus 0.4em\relax Harlow, England: Addison-Wesley, 1999.

% \end{thebibliography}

\bibliographystyle{IEEEtran}

\bibliography{IEEEtran/0tmtt2022}

% biography section
% 
% If you have an EPS/PDF photo (graphicx package needed) extra braces are
% needed around the contents of the optional argument to biography to prevent
% the LaTeX parser from getting confused when it sees the complicated
% \includegraphics command within an optional argument. (You could create
% your own custom macro containing the \includegraphics command to make things
% simpler here.)
%\begin{IEEEbiography}[{\includegraphics[width=1in,height=1.25in,clip,keepaspectratio]{mshell}}]{Michael Shell}
% or if you just want to reserve a space for a photo:

% \begin{IEEEbiography}{Michael Shell}
% Biography text here.
% \end{IEEEbiography}

% % if you will not have a photo at all:
% \begin{IEEEbiographynophoto}{John Doe}
% Biography text here.
% \end{IEEEbiographynophoto}

% % insert where needed to balance the two columns on the last page with
% % biographies
% %\newpage

% \begin{IEEEbiographynophoto}{Jane Doe}
% Biography text here.
% \end{IEEEbiographynophoto}

% You can push biographies down or up by placing
% a \vfill before or after them. The appropriate
% use of \vfill depends on what kind of text is
% on the last page and whether or not the columns
% are being equalized.

%\vfill

% Can be used to pull up biographies so that the bottom of the last one
% is flush with the other column.
%\enlargethispage{-5in}

\include{IEEEtran/Biography}
\end{document}

%% file: Biography.tex
\begin{IEEEbiography}[{\includegraphics[width=1in,height=1.25in,clip,keepaspectratio]{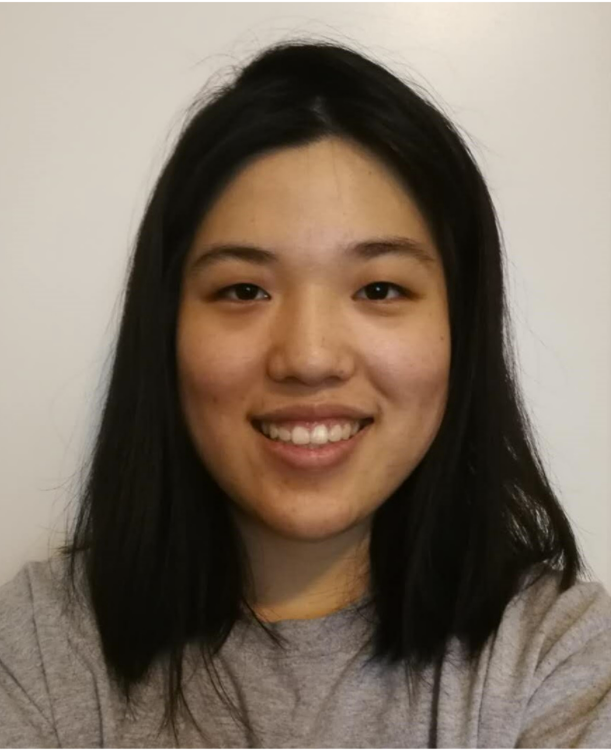}}]
{Jie Yang}
(S`20) received the B.S. degree from Rose-Hulman Institute of Technology, Terre Haute, IN, USA, in 2020. She is currently pursuing the Ph.D. degree in electrical engineering at Purdue University, West Lafayette, IN, USA.
Her research interests include RF systems and circuits. She is also interested in analog and mixed-signal IC design. Ms. Yang is the recipient of CSME Traineeship Fellowship (2021-2022).
\end{IEEEbiography}

\vspace{-0mm}
\begin{IEEEbiography}[{\includegraphics[width=1in,height=1.25in,clip,keepaspectratio]{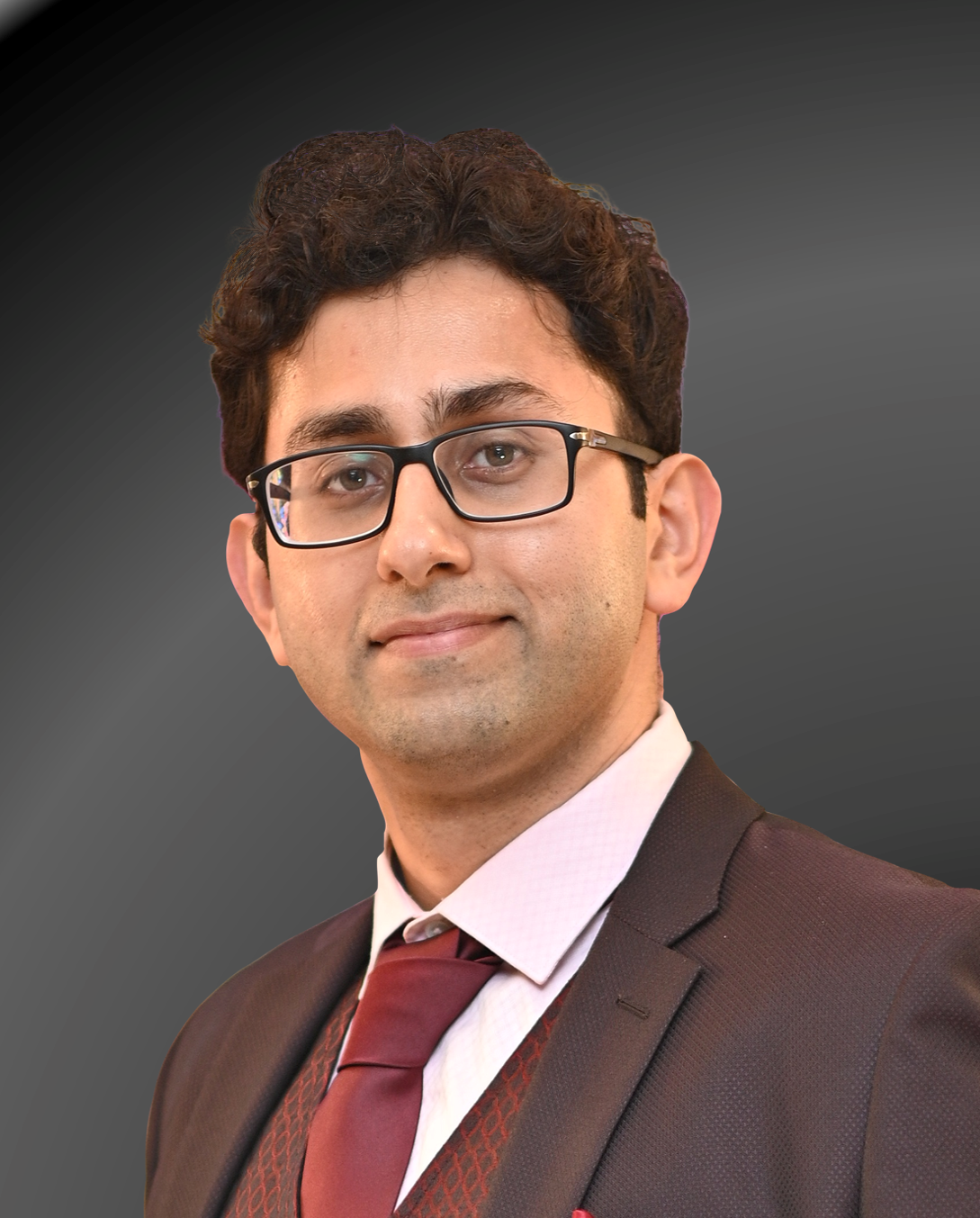}}]
{Baibhab Chatterjee}
 (S`17-M'22) received his Ph.D. from the Elmore Family School of Electrical Engineering, Purdue University, West Lafayette, IN, USA in 2022. He is currently an Assistant Professor in the department of ECE, University of Florida, Gainesville, FL, USA. His industry experience includes two years as a Digital Design Engineer/Senior Digital Design Engineer with Intel, Bengaluru, India, and one year as a Research and Development Engineer with Tejas Networks, Bengaluru, India. He was a Quantum Hardware Design Intern with IBM T.J. Watson Research Center, NY, USA during 2020-2021, where he worked on ultra-low power quantum receiver front-ends.
 
 Dr. Chatterjee was a recipient of the Andrews Fellowship (2017-2019) and the Bilsland Dissertation Fellowship (2021-2022) at Purdue University, the RFIC/IMS 2020 3MT Audience Choice Award, along with 4 best paper/poster awards in HOST 2018, HOST 2019, CICC 2019 and CICC 2021. His research interests include low-power analog, RF, and mixed-signal circuit design for next-generation biomedical, military and quantum applications.
\end{IEEEbiography}

\vspace{-0mm}
\begin{IEEEbiography}[{\includegraphics[width=1in,height=1.25in,clip,keepaspectratio]{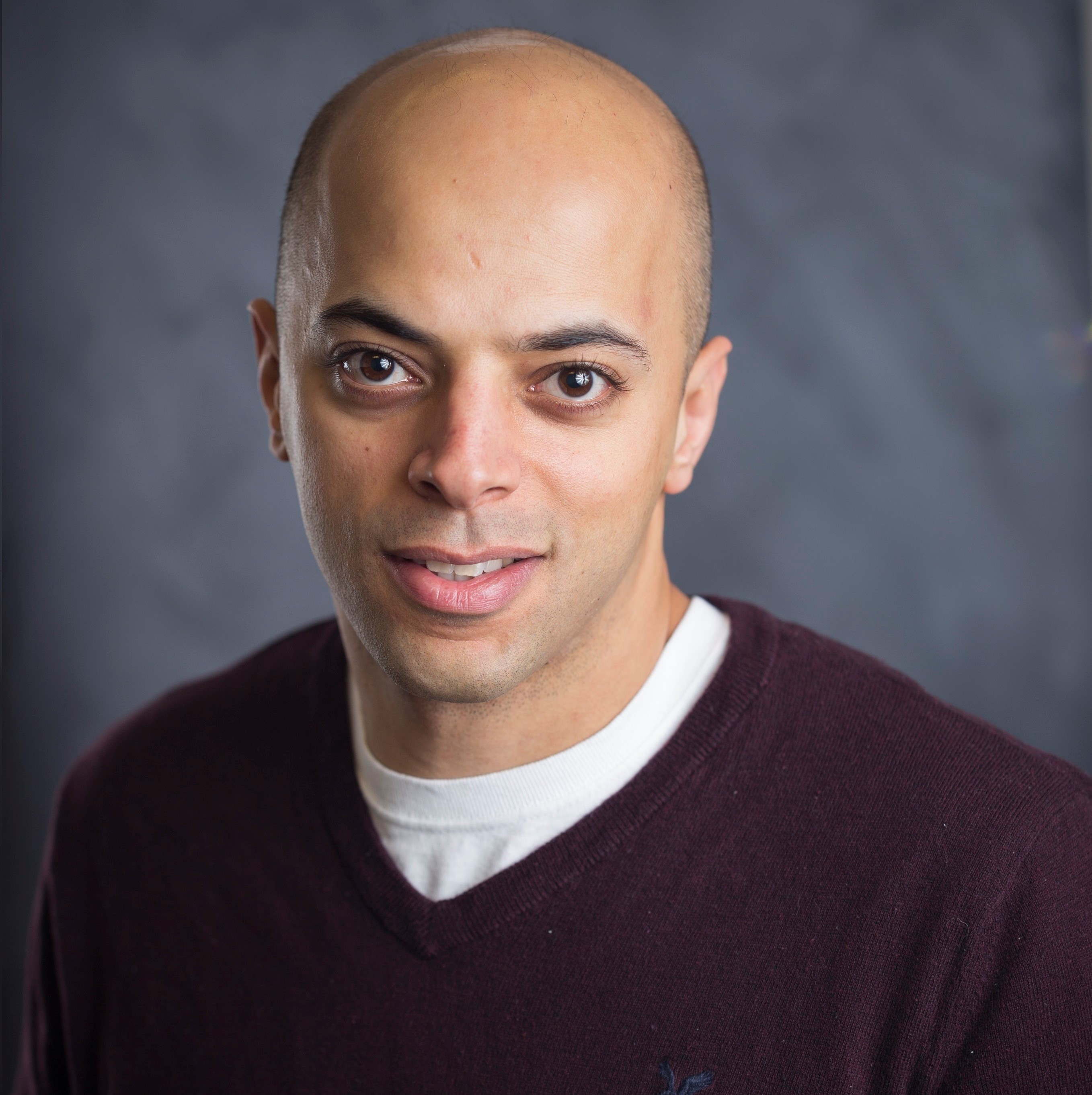}}]
{Mohammad Abu Khater}
(SM’18) received his PhD in Electrical and Computer Engineering from Purdue University, West Lafayette, IN, USA, in 2015. His research interests are primarily focused on adaptive wireless devices, interference detection and suppression, and system-level RF designs. He is currently a research scientist at Purdue University. He was with Qualcomm and Intel Labs where he worked on various high-speed and low-power circuits and systems. He received the Fulbright graduate scholarship. 
Dr. Abu Khater is also a recipient of the best paper award (WAMICON 2022), and a co-recipient of the best paper award from the IEEE Microwave and Wireless Component Letters. He received the excellence in teaching award from the college of engineering at Purdue University in 2012. 
\end{IEEEbiography}

\vspace{-0mm}
\begin{IEEEbiography}[{\includegraphics[width=1in,height=1.25in,clip,keepaspectratio]{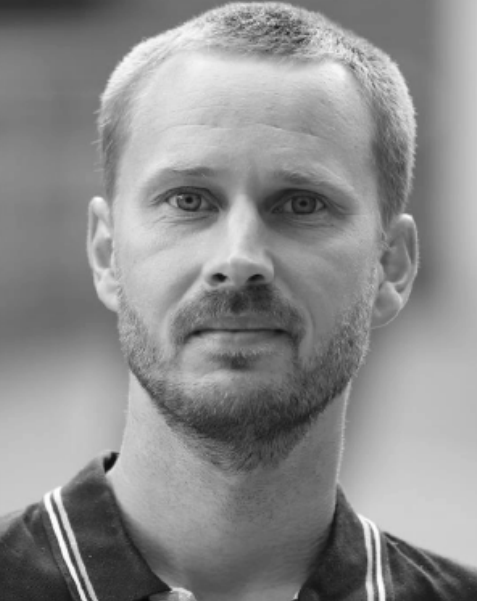}}]
{Mattias Thorsell}
(S’08–M’11) received the M.Sc. and Ph.D. degrees in electrical engineering from Chalmers University of Technology, Gothenburg, Sweden, in 2007 and 2011, respectively. He is currently part time an Associate Professor with Chalmers University of Technology as well as part time Research Leader within Wide Bandgap Technologies at Saab AB. His research interests are electro-thermal characterization and modeling of nonlinear microwave semiconductor devices. 
\end{IEEEbiography}

\vspace{-0mm}
\begin{IEEEbiography}[{\includegraphics[width=1in,height=1.25in,clip,keepaspectratio]{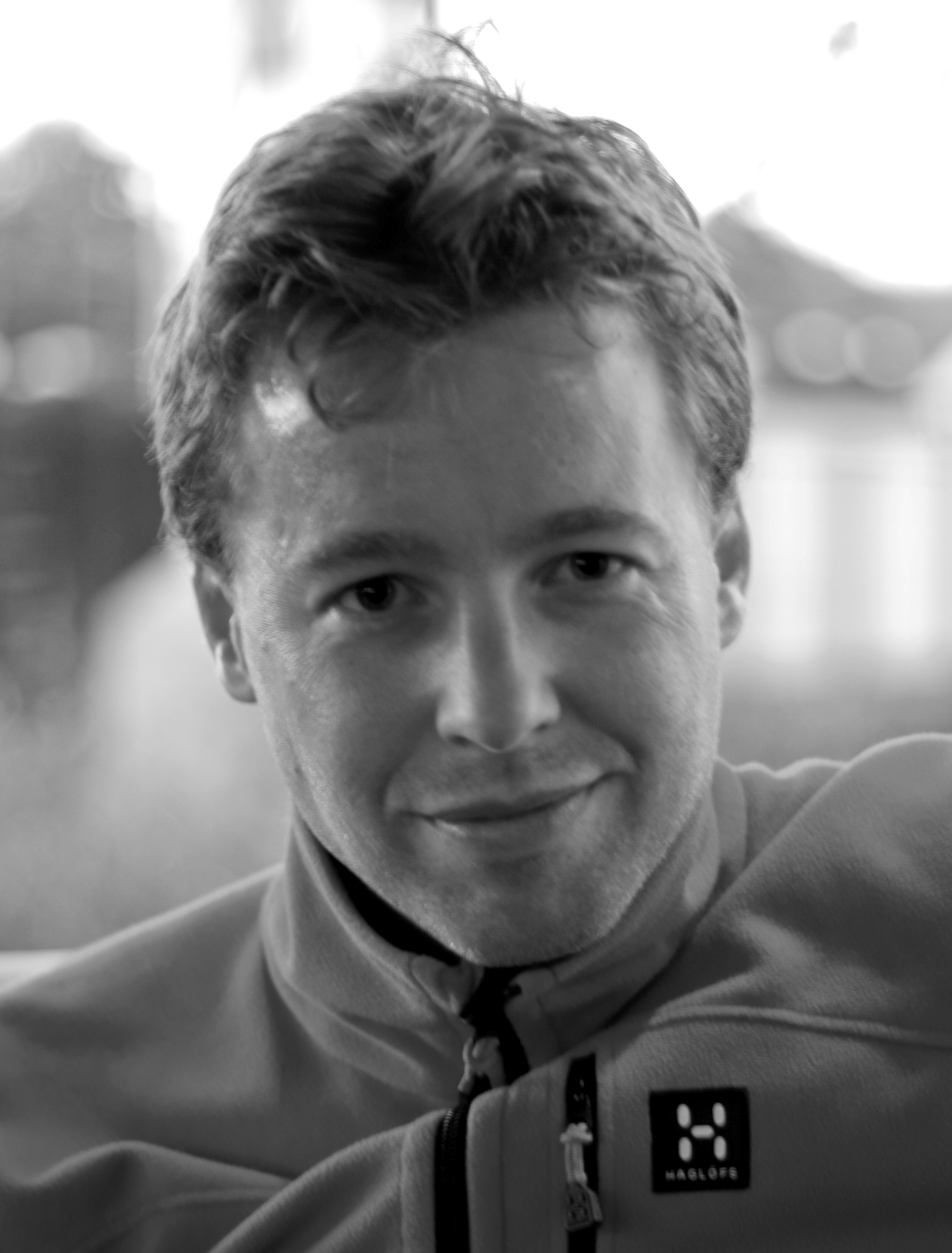}}]
{Sten E. Gunnarsson}
(S’02–M’08-SM’18) received the M.Sc. degree in electrical engineering from the Lund University of Technology, Lund, Sweden, in 2003. He received the Ph.D. degree in mm-wave MMIC design and the Docent degree in Microwave Electronics, both from Chalmers University of Technology, Göteborg, Sweden in 2008 and 2016, respectively. Gunnarsson is currently appointed Specialist within “Microwave Design” at the Microwave and Antenna Group at SAAB AB, Järfälla, Sweden. He is also an Adjunct Professor with the Microwave Electronics Laboratory, Department of Microtechnology and Nanoscience (MC2), Chalmers University of Technology, Sweden, where he is involved in supervision and research. He was previously with Sivers IMA AB where he designed frequency converters and chip-scale packages in the frequency range from 50 to 90 GHz. He is a co-founder of GotMIC AB, a fabless and independent design house of advanced mm-wave MMIC solutions. He has authored or co-authored more than 50 peer-reviewed scientific papers and hold six patents. His main research interest concerns the design of MMICs and packaging solutions for wireless systems operating in the DC to 340 GHz range with focus on extremely high relative bandwidth and/or high operating frequency.
Dr. Gunnarsson was the recipient of the IEEE Microwave Theory and Techniques Society (IEEE MTT-S) Graduate Fellowship Award in 2006 and once again in 2007.
\end{IEEEbiography}

\vspace{-0mm}
\begin{IEEEbiography}[{\includegraphics[width=1in,height=1.25in,clip,keepaspectratio]{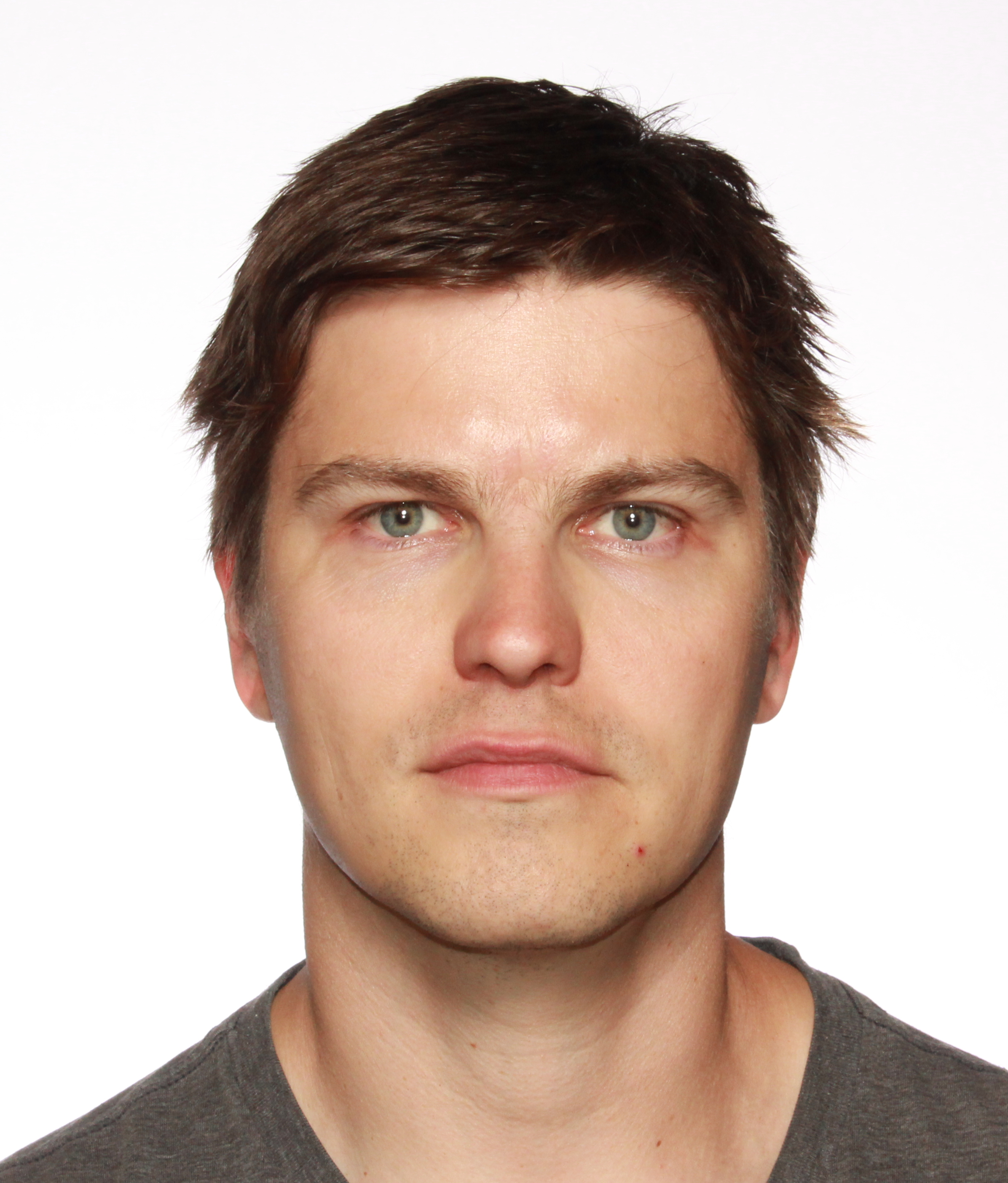}}]
{Tero Kiuru}
received the M.Sc. (Tech.) degree in electrical engineering from the TKK Helsinki University of Technology, Espoo, Finland, and the D.Sc. (Tech.) degree (with distinction) in electrical engineering from Aalto University, Espoo, in 2011.
He is a Research Leader with Saab Finland Oy, and has over 15 years of experience in the research activities in high-frequency radio technology, radar technology and electronic warfare systems. He has worked at the Department of Radio Science and Engineering, Aalto University, the European Space Agency’s Reseach Centre ESTEC in Noordwijk, The Netherlands, NASA’s Jet Propulsion Laboratory, Pasadena, CA, USA, and at VTT Technical Research Centre of Finland and Saab Finland Oy. 
He has published more than 40 scientiﬁc articles in the ﬁeld of microwave components and systems, radar technologies and high-frequency measurements. His current research interests include novel microwave radar and electronic warfare systems.
\end{IEEEbiography}

\vspace{-0mm}
\begin{IEEEbiography}[{\includegraphics[width=1in,height=1.25in,clip,keepaspectratio]{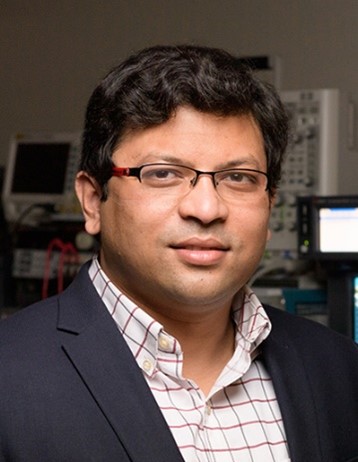}}]
{Shreyas Sen} (S`06-M`11-SM`17) received the Ph.D. degree in ECE from Georgia Tech, Atlanta, GA, USA, in 2011. He has over five years of industry research experience in Intel Labs, Hillsboro, OR, USA, Qualcomm, Austin, TX, USA, and Rambus, Los Altos, CA, USA. He is currently an Elmore Associate Professor of ECE \& BME, Purdue University, West Lafayette, IN, USA. Dr. Sen is the inventor of the Electro-Quasistatic Human Body Communication (EQS-HBC), or Body as a Wire Technology, for
which, he was a recipient of the MIT Technology Review top-ten Indian Inventor Worldwide under 35 (MIT TR35 India) Award. His work has been covered by 250+ news releases worldwide, invited appearance on TEDx Indianapolis, Indian National Television CNBC TV18 Young Turks Program, NPR subsidiary Lakeshore Public Radio, and the CyberWire podcast. He has authored/coauthored three book chapters, over 175 journal and conference papers, and has 15 patents granted/pending. He serves as the Director for the Center for Internet of Bodies (C-IoB). His current research interests span mixed-signal circuits/systems and electromagnetics for the Internet of Things (IoT), biomedical, and security.

Dr. Sen is a recipient of the NSF CAREER Award in 2020, the AFOSR
Young Investigator Award in 2016, the NSF CISE CRII Award in 2017, the Intel Outstanding Researcher Award in 2020, the Google Faculty Research Award in 2017, the Purdue CoE Early Career Research Award in 2021, the Intel Labs Quality Award in 2012 for industry-wide impact on USB-C type, the Intel Ph.D. Fellowship in 2010, the IEEE Microwave Fellowship in 2008, the GSRC Margarida Jacome Best Research Award in 2007, and nine best paper awards, including IEEE CICC in 2019, 2021, and in IEEE HOST 2017–2020, for four consecutive years. His work was chosen as one of the top-ten papers in the hardware security field (TopPicks 2019). He serves/has served as an Associate Editor for IEEE SOLID STATE CIRCUITS LETTERS (SSC-L), \textit{Frontiers in Electronics}, \textit{IEEE Design \& Test}, an Executive Committee Member of the IEEE Central Indiana Section and a Technical Program Committee Member of the ACM Design Automation Conference (DAC), the IEEE Custom Integrated Circuit Conference (CICC), Design, Automation and Test in Europe (DATE), the ACM/IEEE International Symposium on Low Power Electronics and Design (ISLPED), the International Conference on Computer-Aided Design (ICCAD), the International Test Conference (ITC), VLSI Design, among others. 
\end{IEEEbiography}